\documentclass{aa}

\usepackage{hyperref}

\usepackage{xcolor}
\definecolor{dark-red}{rgb}{0.9,0.0,0.0}
\definecolor{dark-blue}{rgb}{0.15,0.15,0.9}
\definecolor{dark-green}{rgb}{0.15,0.8,0.15}
\definecolor{medium-blue}{rgb}{0,0,0.9}
\hypersetup{
    colorlinks, linkcolor=red,
    citecolor={dark-blue} , urlcolor={medium-blue}
}
\usepackage{changepage} % for \begin{adjustwidth}{-Xcm}{} \end{adjustwidth}

\usepackage{graphicx}
\usepackage{txfonts}
\begin{document} 

   \title{Precise radial velocities of giant stars \\
   }
   \subtitle{VIII. Testing for the presence of planets with CRIRES Infrared Radial Velocities
      \thanks{Based on observations collected at the European Southern Observatory, 
Chile, under program IDs 088.D-0132, 089.D-0186, 090.D-0155 and 
091.D-0365.}}

    \author{Trifon Trifonov\inst{1}$^{,}$\inst{2}, Sabine Reffert\inst{1}, Mathias Zechmeister\inst{3}, Ansgar Reiners\inst{3}, 
    \and Andreas Quirrenbach\inst{1}}
 
   \institute{Landessternwarte, Zentrum f\"{u}r Astronomie der Universit\"{a}t Heidelberg,\ K\"{o}nigstuhl 12, 69117 Heidelberg, Germany\\ \vspace{-3mm}
              \and
              Department of Earth Sciences, The University of Hong Kong, Pokfulam Road, Hong Kong \\ \vspace{-3mm}
              \and
              Institut f\"{u}r Astrophysik, Georg-August-Universität, Friedrich-Hund-Platz 1, 37077 G\"{o}ttingen, Germany
 \\}
   
   \date{Received March 27, 2015; accepted August 4, 2015}

% \abstract{}{}{}{}{} 
% 5 {} token are mandatory
 
  \abstract
   {We have been monitoring 373 very bright (V $\leq$ 6 mag) G and K giants
 with high precision optical Doppler spectroscopy for more than a decade at Lick Observatory.
 Our goal was to discover planetary companions around those stars and 
 to better understand planet formation and evolution around intermediate-mass stars. 
  However, in principle, long-term, $g$-mode nonradial stellar pulsations or rotating 
  stellar features, such as spots, could effectively mimic a planetary signal in the radial velocity data. }
  {Our goal is to compare optical and infrared radial velocities for those stars with periodic radial 
   velocity patterns and to test for
   consistency  of their fitted radial velocity semiamplitudes. Thereby, we distinguish
   processes intrinsic to the star 
   from orbiting companions as reason for the radial velocity periodicity observed in the optical.}
   {Stellar spectra with high spectral resolution have been taken in the \textit{H}-band with the 
   CRIRES near-infrared spectrograph at ESO's VLT for 20 stars of our Lick survey. 
    Radial velocities are derived using 
    many deep and stable telluric CO$_2$ lines for precise wavelength calibration.}
   {We find that the optical and  near-infrared radial velocities of the giant stars in our sample are 
    consistent. We present detailed results for eight stars in our sample previously reported to have planets or
    brown dwarf companions.
    All eight stars passed the infrared test.}
    {We conclude that the
    planet hypothesis provides the best explanation for the periodic radial velocity patterns
    observed for these giant stars.}

      \authorrunning{Trifon Trifonov et al.}
   \maketitle

\section{Introduction}

By June 2015, the number of confirmed substellar companions discovered with the 
Doppler technique has reached $\sim$ 600\footnote{http://exoplanet.eu}.
This number constitutes $\sim$ 30\% of the total number of planets discovered using all planet search methods.
In addition, many more extrasolar planets, in particular those discovered with
the transiting method, have been confirmed with Doppler spectroscopy.
This shows that the precise radial velocity (RV) method still remains one of the most valuable
extrasolar planet search tools to date. 

Ultra stable  \'{e}chelle spectrographs, 
such as HARPS \citep{Mayor2003}, have already reached sub-m\,s$^{-1}$ precision 
and potentially allow astronomers to search for Earth mass planets.
The Doppler technique, however, can suffer from false positive detections due to 
radial velocity variations intrinsic to a star. 
In addition to short period $p$-mode pulsations \citep{Barban, Zechmeister2008}, which are excited by 
convection and seen as stellar radial velocity scatter, 
stellar surface spots or even nonradial gravitational $g$-mode pulsations can potentially 
cause line shape deformations, which can be misinterpreted as velocity shifts.

So far, more than 60 substellar companions have been detected around evolved G and K giant stars
using the Doppler method. 
Some of these stars have RV signals clearly consistent with highly eccentric substellar companions
that cannot be mistaken for stellar activity \citep[e.g.,][]{Frink2,Moutou2011,Sato2013},
but for others an alternative explanation of the data cannot be excluded despite their periodic RV signals.
While long-term $g$-mode pulsations are rather unlikely to be excited  
in the large convective layers in these evolved stars,  
temperature spots on the other hand could effectively mimic a planet \citep[e.g.,][]{Hatzes2000}. 
As K~giants are inflated stars compared to their main-sequence progenitors, 
their rotational period can be of the order of hundreds of days.
Therefore, in case of spot(s) on the stellar surface, the line profile variations can lead to 
RV variations with long periods and semiamplitudes significantly surpassing the intrinsic stellar jitter
that can be easily mistaken for a planet.

Detailed studies of spectral line profile bisectors have often been used to support the companion hypothesis.
This method requires very high resolution spectra, however, and a stable instrumental profile, and the absence of variations in the spectral line shape is  a necessary condition, 
but is not sufficient to  prove the existence of a substellar companion definitively.

Measuring RV signals in the near-infrared (near-IR) is another promising test that can be performed. The contrast 
ratio between the star's surface blackbody intensity and that of a cooler spot 
is much larger in the optical than in the IR domain. In case of spots, 
 line shape deformations in the IR data must have a much smaller amplitude \citep{Desort2007, Reiners2010}. 
We can also expect different RV amplitudes in the two domains 
in case of nonradial $g$-mode pulsations, while this test should yield no difference 
in the optical and near-IR radial velocities in case of a planet.

A prominent example for the discussion of spot-induced radial velocities is the young star TW Hya; \citet{Setiawan2008} found 
periodic radial velocity variations, but no indication of line bisector variations and concluded that 
the star is orbited by a planet. 
Later, \citet{Huelamo} observed TW Hya at infrared wavelengths but could 
not find RV variations consistent with the Keplerian solution from 
optical measurements. \citet{Huelamo} concluded that no planet orbits TW Hya.

High precision RVs derived from the infrared wavelength regime can form a rather critical test for planet confirmation. 
This test is valid mostly for active stars for which other viable explanations for periodic 
radial velocities, such as spots or pulsations, cannot be excluded by any other means, such as giant stars in particular.

An obvious choice for this test is the ESO pre-dispersed CRyogenic InfraRed Echelle Spectrograph (CRIRES),
mounted at the Nasmyth focus B of the 8m VLT UT1 \citep{Kaeufl}.
Several studies  demonstrated that radial velocity measurements with precision 
between 10 and 35~m\,s$^{-1}$ are possible with CRIRES.
\citet{Seifahrt} reached a precision of $\approx$~35~m\,s$^{-1}$ for CRIRES radial velocities, using the 
N$_2$O gas cell for calibration. \citet{Huelamo} and \citet{Figueira} showed that a Doppler precision of 
$\approx$~10~m\,s$^{-1}$ can be achieved when using telluric CO$_2$ lines in the $H-$band as wavelength reference.

We present our CRIRES near-infrared Doppler results for 20 evolved G8-K4\,III~giant stars.
All selected targets in this study have periodic radial velocities as derived from optical spectra, and are thus potential planet hosts. 
They constitute a small subsample of our K~giant planet search sample in the optical \citep{Frink, Reffert2014}.

In Section~\ref{The Kgiant sample} we briefly introduce the background of this program.  
Section~\ref{Observational setup} describes our observational setup with CRIRES. 
In Section~\ref{Calibration and data reduction} we explain in detail the data reduction and analysis process with our CRIRES pipeline.
Our methods of extracting the precise RVs are given in Section~\ref{Obtaining the radial velocities}, and in Section~\ref{IR data analysis}
we discuss the consistency between optical and IR data. In Section~\ref{Results and Disc} we 
discuss our results, and we provide a summary in Section~\ref{Summary}.

\begin{table*}{}      
\centering    

\caption{List of observed stars.}   
\label{table:List} 

\resizebox{0.75\textheight}{!}{\begin{minipage}{\textwidth}
\centering

 \begin{tabular}{llcccr @{ }lccc}
\hline
\noalign{\vskip 0.5mm}
\hline
\noalign{\vskip 0.9mm}

HIP number &HD number &Sp.type$^{\alpha}$&$M^{\beta}$ $[M_{\odot}]$& $R^{\beta}$ $[R_{\odot}]$& $L^{\beta}$& $[L_{\odot}]$&
$V^{\alpha}$ [mag] & $H^{\gamma}$ [mag]& $N$ obs.\\  
\hline
\noalign{\vskip 0.9mm}

 5364    &  6805    & K2\,III & 1.7 $\pm$ 0.1 & 14.3  $\pm$ 0.2  & 71.1   & $\pm$ 1.1     & 3.46 & 1.0 & 9  \\
 19011   &  25723   & K1\,III & 2.1 $\pm$ 0.3 & 13.9  $\pm$ 0.6  & 89.5   & $\pm$ 7.2     & 5.62 & 3.1 & 6  \\
 20889   &  28305   & K0\,III & 2.4 $\pm$ 0.2 & 13.1  $\pm$ 0.2  & 85.3   & $\pm$ 2.7     & 3.53 & 1.3 & 8  \\
 23015   &  31398   & K3\,II~ & 4.5 $\pm$ 0.6 & 126.3 $\pm$ 12.6 & 3752.8 & $\pm$ 675.1     & 2.69 & $-$0.7~~ & 6 \\
 31592   &  47205   & K1\,III & 1.4 $\pm$ 0.2 & ~~5.1 $\pm$ 0.3  & 12.0   & $\pm$ 0.6     & 3.95 & 1.7 & 9  \\
 34693   &  54719   & K2\,III & 2.3 $\pm$ 0.3 & 36.8  $\pm$ 0.7  & 238.4  & $\pm$ 11.4    & 4.41 & 1.8 & 8  \\
 36616   &  59686   & K2\,III & 1.9 $\pm$ 0.2 & 13.2  $\pm$ 0.3  & 73.3   & $\pm$ 3.4     & 5.45 & 3.1 & 8  \\
 37826   &  62509   & K0\,III & 2.3 $\pm$ 0.2 & ~~8.7 $\pm$ 0.3  & 39.8   & $\pm$ 0.6     & 1.16 & $-$0.8~~ & 8  \\
 38253   &  63752   & K3\,III & 2.0 $\pm$ 0.5 & ~~69.4  $\pm$ 11.3 & 1169.2 & $\pm$ 379.8 & 5.60 & 2.5 & 8  \\
 39177   &  65759   & K3\,III & 2.7 $\pm$ 0.6 & 20.4  $\pm$ 1.4  & 473.2  & $\pm$ 134.2   & 5.60 & 2.9 & 7  \\
 60202   &  107383  & G8\,III & 2.2 $\pm$ 0.3 & 15.6  $\pm$ 0.4  & 112.0  & $\pm$ 4.6     & 4.72 & 2.5 & 7  \\
 73133   &  131918  & K4\,III & 1.2 $\pm$ 0.1 & 44.0  $\pm$ 2.1  & 408.2  & $\pm$ 39.8    & 5.48 & 2.3 & 6  \\
 74732   &  135534  & K2\,III & 1.5 $\pm$ 0.2 & 26.5  $\pm$ 1.1  & 200.1  & $\pm$ 16.0    & 5.52 & 2.7 & 7  \\
 79540   &  145897  & K3\,III & 1.1 $\pm$ 0.2 & 26.8  $\pm$ 0.8  & 190.9  & $\pm$ 10.5    & 5.24 & 2.3 & 8  \\
 80693   &  148513  & K4\,III & 1.2 $\pm$ 0.1 & 29.5  $\pm$ 1.6  & 206.4  & $\pm$ 21.5    & 5.41 & 2.3 & 8  \\
 84671   &  156681  & K4\,II~   & 1.2 $\pm$ 0.1 & 53.4  $\pm$ 2.7  & 554.0  & $\pm$ 54.4    & 5.03 & 1.5 & 5  \\
 88048   &  163917  & K0\,III & 2.7 $\pm$ 0.2 & 14.6  $\pm$ 0.3  & 109.3  & $\pm$ 3.1     & 3.32 & 1.3 & 10  \\
 91004   &  171115  & K3\,III & 6.4 $\pm$ 2.0 & \ldots      & \multicolumn{2}{c}{\ldots}  & 5.49 & 1.3 & 7  \\
 100587  &  194317  & K3\,III & 1.4 $\pm$ 0.2 & 23.6  $\pm$ 0.5  & 159.3  & $\pm$ 5.4     & 4.43 & 1.4 & 7  \\
 114855  &  219449  & K0\,III & 1.4 $\pm$ 0.1 & 11.0 $\pm$ 0.1   & 51.5   & $\pm$ 1.1     & 4.24 & 1.7 & 8  \\
\hline
\noalign{\vskip 0.5mm}
%\hline
%\noalign{\vskip 0.5mm}
 \makebox[0.1\textwidth][l]{$\alpha$ - Hipparcos Catalog, $\beta$ - \citet{Reffert2014}, $\gamma$ - 2MASS Catalog}\par \\

\end{tabular}
%\end{adjustwidth}
\end{minipage}}
\end{table*}

%__________________________________________________________________
\section{The K~giant sample}
\label{The Kgiant sample}

Our Doppler survey started in 1999 at Lick Observatory using
the Hamilton spectrograph in conjunction with an iodine cell \citep{Marcy2,Butler}.
The initial goal of our program and the star selection criteria are described in \citet{Frink}.
Briefly, we regularly observed 373 very bright (V~$\leq$~6~mag) and photometrically constant G and K~giants selected from the Hipparcos Catalog.
The objective of our program is to investigate and understand giant planet occurrence and evolution around intermediate-mass stars. 

The first planet from our program was discovered around the K~giant star $\iota$~Dra \citep{Frink2}. 
The highly eccentric Keplerian reflex motion of $\iota$~Dra seen in the Lick data leaves no doubt on the planet's existence. 
Therefore, $\iota$~Dra~b became the first confirmed extrasolar planet around a giant star and encouraged
more scientists to search for extrasolar planets around evolved stars with the Doppler method. As a result, 
 to date more than 60 planets\footnote{http://www.lsw.uni-heidelberg.de/users/sreffert/giantplanets.html}
have been discovered around giants, and their number is constantly growing.

Planet occurrence statistics and more results from our Lick survey are given in \citet{Reffert2014}.
We distinguish between planets and planet candidates among the companions,
depending on how persistent the RV signal is over many cycles, and how large
the RV amplitude is compared to the intrinsic RV jitter caused by short-term radial pulsations.

A subsample of 20~stars with planets and planet candidates from our survey is 
of particular interest, and are studied further here. 
All these stars display either one or two clear periodicities in their RVs,
consistent with one or two orbiting substellar companions with minimum masses ranging between
a few Jupiter masses and a low-mass brown dwarf.
If stellar spots were responsible for the obtained Doppler velocities, they would correspond 
to photometric variability \citep{Hatzes2002}.
Precise photometry from Hipparcos shows that our sample is photometrically
stable down to 3~mmag, and thus there are no indications for any intrinsic 
features to be related to a Doppler signal \citep{Reffert2014}.
Nevertheless, since we cannot fully exclude long-period pulsations 
(\textit{g}$-$mode or of unknown nature), we have observed our subsample of 20~stars with CRIRES over four semesters
in an attempt to provide more evidence in favor of the companion hypothesis.
Stars are listed in Table~\ref{table:List}, including their Hipparcos and HD number, apparent magnitude in $V$ and $H$ bands,
as well as basic physical parameters, such as stellar mass $M$, radius $R,$ and luminosity $L$ as  given in \citet{Reffert2014}.
Some planetary systems from the subsample have already been published: 
HIP~37826  \citep{Reffert}, HIP~88048  \citep{Quirrenbach}, HIP~34693 and HIP~114855 \citep{Mitchell2013}, and HIP 5364  \citep{Trifonov2014}.
 Others have been discovered independently: 
HIP~37826  \citep{Hatzes}, HIP~20889  \citep{Sato2007}, HIP~60202  \citep{Liu2008}, and HIP~31592  \citep{Wittenmyer2011}.
 Publication of the remaining stars with confirmed planets is in preparation.

\section{Observational setup}
\label{Observational setup}

%CRIRES\LEt{Please avoid beginning sentences with abbreviations, acronyms, numbers in figures, and the like. Please check for this throughout the paper.  .} is a temperature stabilized near-infrared spectrograph with 
%CRIRES is a temperature stabilized near-infrared spectrograph with a
%The CRyogenic InfraRed Echelle Spectrograph (CRIRES) is a temperature stabilized instrument with 
The ESO's CRIRES instrument is a temperature stabilized near-infrared spectrograph with a
maximum resolution of $R$~$\approx$~100\,000 when used with a 0.2\arcsec~slit.
Despite a broad accessible wavelength range from 950 to 5200 nm (J,~H,~K,~L, and M infrared bands),
the available wavelength coverage per single observation in this predispersed 
\'{e}chelle spectrograph is much smaller than in optical cross-dispersed 
\'{e}chelle spectrographs and currently limited to one spectral order\footnote{Currently, CRIRES is removed from UT1 for an upgrade and is expected to be operational 
again in 2017 under the name CRIRES+. The new instrument is expected to
cover simultaneously a wavelength range about ten times larger than the original \citep{Dorn2014}.}.
A particular wavelength setting can be selected for each observation, with a free spectral
range on the order of 20-200 nm, depending on the spectral order and the dispersion for that setting. 
The spectrum is imaged by a mosaic of four Aladdin III detectors, forming an effective 4096~x~512 pixel array 
(with a gap of $\sim$ 280 px between the detectors). 
This arrangement is somewhat problematic because a precise wavelength solution must be obtained separately for each detector. 
The standard CRIRES pipeline recipes offer in principle this kind of  calibration, but at a level of radial velocity precision 
far below that required for the Doppler detection of planets. 

Thus, we had two options when calibrating our spectra: to use a similar approach to 
the I$_2$ cell method  using N$_2$O or NH$_3$ gas cells \citep{Seifahrt, Bean2}, or to use atmospheric telluric lines \citep{Huelamo,Figueira}.
We chose to follow the telluric method.

\subsection{Spectral window}
\label{Spectral window}
 
Our wavelength setting was selected by inspecting the Arcturus near-IR spectral atlas \citep{Hinkle1995}. 
We searched for dense stellar line regions that also contained deep and sharp telluric lines to be used for wavelength calibration. 

We chose a wavelength setting in the $H$-band (36/1/n in the CRIRES user manual), 
with a reference wavelength of $\lambda$ = 1594.5~nm (wavelength in the middle of detector 3). 
In fact, the selected region is very close to the wavelength setup 
successfully used by \citet{Figueira}. Our setup is characterized by the presence 
of many sharp atmospheric CO$_2$ lines, which we used as wavelength calibrators. 
Unfortunately, the CO$_2$ lines only fall  on detector 1 and 4.
Although detectors 2 and 3 contain many stellar lines, we were not able to construct
a wavelength solution for these spectra on these detectors, and thus we excluded 
detectors 2 and 3 from our analysis.

\subsection{Observations}
\label{ObservationsCRIRES}

We performed the observations {}with a standard ABBA nodding cycle, including three jitter observations per nod 
 to subtract the sky background emission. The total exposure time required for each individual target to 
reach a S/N $\geq$ 300 at the reference wavelength has been estimated with the ESO Exposure Time 
Calculator\footnote{http://etimecalret.eso.org/observing/etc/bin/gen/form?INS.NAME= CRIRES+INS.MODE=swspectr} (ETC) 
to be between 18 seconds for the brightest stars and 3 minutes for the faintest stars. To achieve the highest 
possible precision, the spectrograph is used with the 0.2\arcsec~slit, resulting in a resolution of $R$ = 100\,000.
To minimize RV errors related to the imperfect stability of the slit illumination,
the observations were done in NoAO mode (without adaptive optics), and nights with poor seeing conditions were requested if possible.

Since the periods derived from optical RV data are typically one to two years long, 
we took at least four RV measurements per year, if possible.
During the four semesters of observations some targets were given higher
priority than others; the number of data points thus ranges from  six to ten.

\section{Calibration and data reduction}
\label{Calibration and data reduction}
 
Dark, flat, and nonlinearity corrections, as well as the combination of raw jittered 
frames in each nodding position, were performed using the standard ESO CRIRES pipeline recipes. 
The final output from the CRIRES pipeline is an extracted, one-dimensional coadded spectrum, 
but we also obtained spectra from the individual A and B nodding frames separately.
The CRIRES pipeline wavelength calibration is based on a robust cross-correlation technique between observed spectra and
ThAr lamp lines or, alternatively, from a physical model employing ray tracing.
However, the precision of both wavelength solutions was insufficient for obtaining precise radial velocities. 
We found the wavelength solution from the physical model to be more precise, and thus we used it as an initial
guess for the calibration of each spectral frame with telluric lines.

\begin{figure*}[htp!]
\begin{center}$
\begin{array}{cc} 
\includegraphics[width=3.55in]{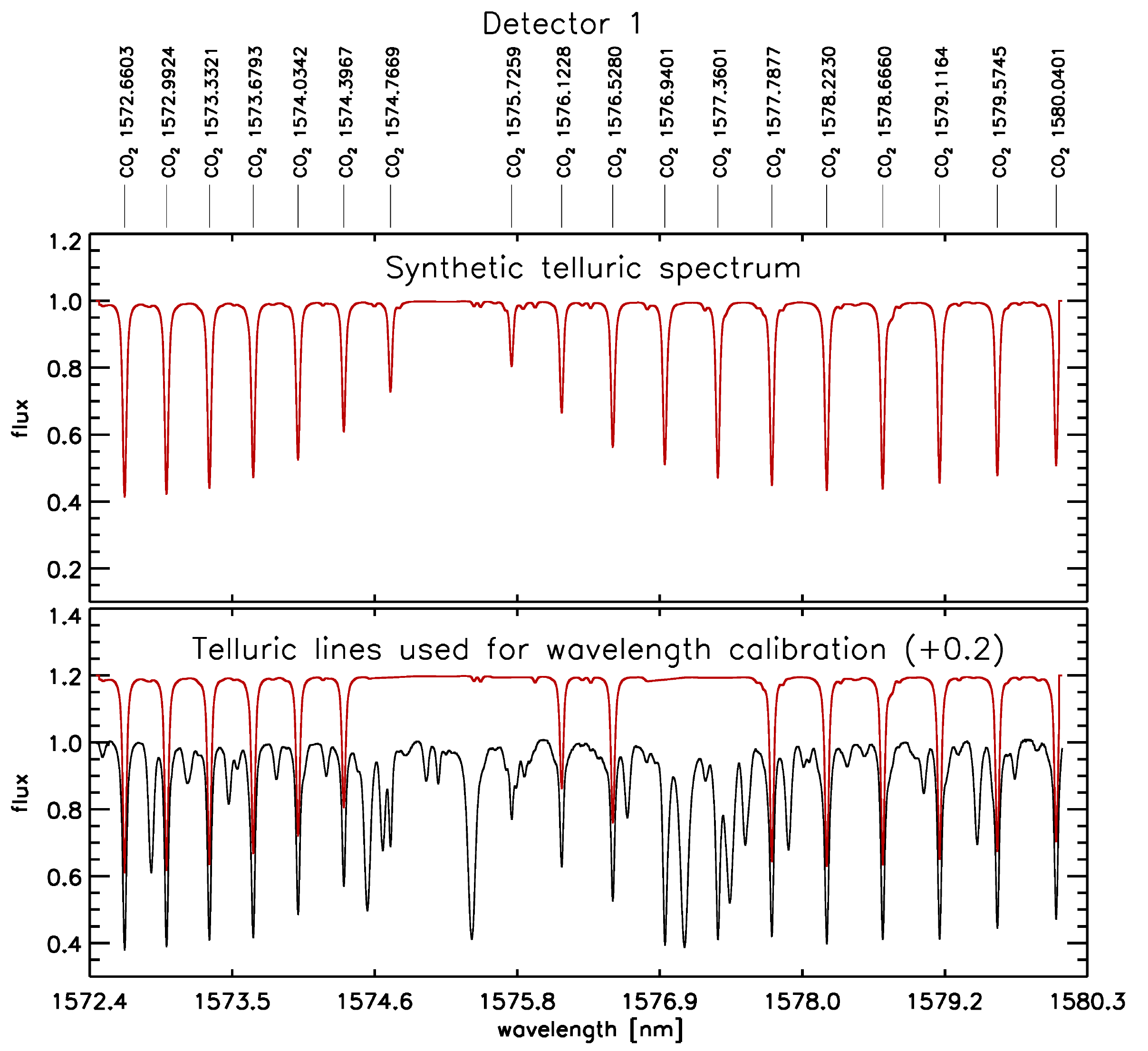} &
\includegraphics[width=3.55in]{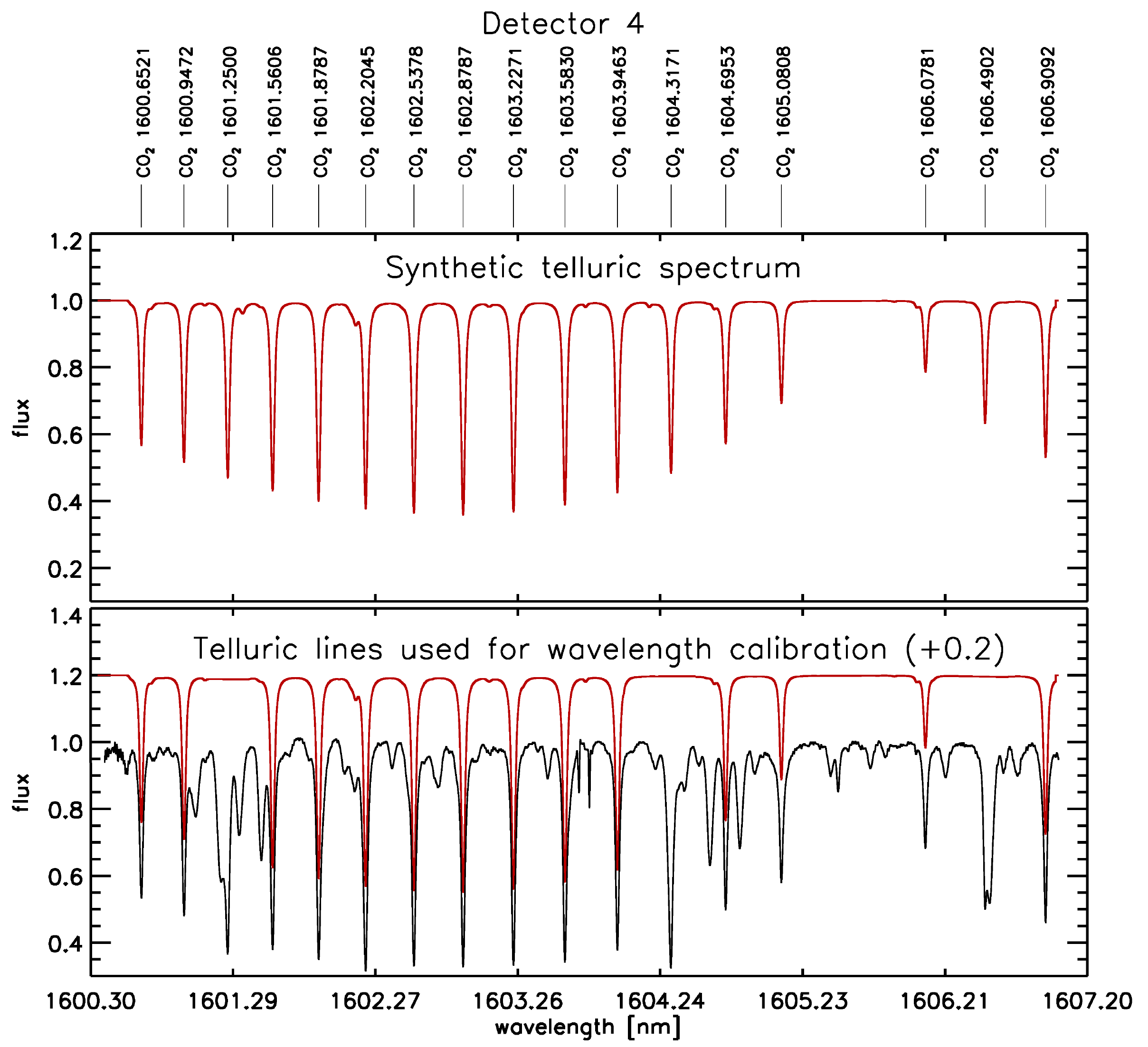} \\
\end{array} $

\end{center}
\caption{Top: model spectra of the telluric transmission for detectors one (left) and four (right).
Bottom: calibrated science spectra for both detectors (black) showing the stellar and telluric absorption lines. 
Only nonblended telluric lines were modeled (red) and used for constructing the wavelength solution.} 
 
\label{FigGam:chip1-4b-cascade}  
\end{figure*}

For further data analysis, we developed a semiautomatic pipeline based on a sequence of 
$\chi^2$ minimization algorithms. Fig.~\ref{FigGam:chip1-4b-cascade} and \ref{FigGam:chip1-4c-cascade}
illustrate our data reduction steps on detector 1 and 4 for the first observation of the K~giant HIP~60202.

\subsection{Normalization}
\label{Normalization and line identification}

Continuum normalization of G and K giant spectra is complicated by the high density of spectral lines,
which leave almost no continuum level points to work with.
To perform a proper continuum normalization, we thus developed an automated algorithm, which identifies areas of the
raw spectrum that are free of absorption lines. To find the real continuum points on the raw spectrum,
the same algorithm was applied on a noise-free spectrum constructed by combining a
theoretical telluric spectrum and a synthetic stellar spectrum, 
which was shifted to correct for the barycentric velocity
valid for the given epoch of observations (see Section~\ref{Wavelength calibration}). 
The matching continuum points between both spectra were selected for the construction of a low-order polynomial 
fit that represents the estimated continuum level. 
The raw individual spectrum was then divided by the continuum solution, yielding a normalized spectrum.
Extensive testing showed that this algorithm performs a much better continuum normalization
in the case of high spectral line densities than the usual algorithm, which assumes that all
points below a given threshold belong to the continuum.

\subsection{Wavelength calibration}
\label{Wavelength calibration}

For precise Doppler measurements with CRIRES, a precise wavelength calibration for each epoch and detector 
must be obtained. We applied several automated steps  to first identify all CO$_2$ lines 
used for wavelength calibration in the spectra and then to derive their line centroids in pixel and wavelength space.
 
First we identified all significant absorption lines in the normalized spectrum 
with relative depths below 0.85 of the continuum level \citep{Pepe}.
Telluric lines from CO$_2$ are nearly static, with relatively equidistant wavelengths, and 
thus were easily identified in the spectra as such. 
Nevertheless, for precise wavelength calibration on the basis of telluric CO$_2$ lines,
one should take into account that the line centroids may vary depending 
on the ambient climate above the observatory at the time when the spectrum is obtained. 
Thus, to assure precise calibration, we adopted methods for telluric spectral synthesis 
similar to those used in \citet{Seifahrt2008}, \citet{Seifahrt2010} and \citet{Lebzelter2012}.

We used the physical wavelength solution from the CRIRES recipes as initial guess  to roughly
identify the wavelength region of each observational epoch and for each detector.
Next, we constructed a high resolution telluric transmission spectrum using the
Line By Line Radiative Transfer Model\footnote{LBLRTM $-$ part of FASCODE \citep{Clough1981,Clough1992}, available at http://rtweb.aer.com/} 
(LBLRTM) code for the wavelength region of interest.
The precise molecular line positions needed for the model were taken from the HITRAN database \citep{Rothman}.
For the atmospheric profile above VLT, we adopted the mid-latitude summer/winter meteorological model, which is implemented in the LBLRTM. 

Additional input parameters for the atmospheric model included the target's zenith angle
as well as ambient atmospheric pressure and temperature at the time of observation. For lack of
anything else we used ground-layer pressure and temperature, which might not adequately reflect
the conditions higher up in the atmosphere.
We were only interested  in the CO$_2$ absorption spectra,
but  to avoid confusion resulting from any unresolved lines that might appear on science spectra
we used all the available atmospheric molecules from HITRAN and obtained the telluric mask. 
Many other very weak lines caused by other molecules appeared in the theoretical telluric spectra, 
but they could not be identified separately in the science spectra.
Weak molecular lines affect the total continuum level, but apart from that we consider their contribution negligible.

The resulting spectra for detectors one and four were always dominated by many deep and sharp CO$_2$ 
lines in absorption. Detector two also showed a few CO$_2$ absorption lines, 
but their intensity declined fast toward longer wavelengths and, in general, the lines were heavily 
blended with stellar lines from the science spectrum. No deep telluric lines could be seen on detector three.

We did not use all of the identified telluric and stellar lines  in the following steps. 
The implemented line identification algorithm effectively excludes lines that are either blended or in close proximity to each other.
The selected lines depend on the stellar spectrum shift, which is caused mostly by Earth's barycentric movement.
For this reason, a different set of telluric lines was used for wavelength calibration in each observational epoch.

In the next step, the unblended observed telluric lines in the science spectrum are interpolated and oversampled with a spline function. 
Line centers are obtained in pixel coordinates by fitting a Gaussian, Lorentzian, or Pseudo-Voigt 
(weighted sum between a Gaussian and a Lorentzian) profile via $\chi^2$ minimization.
The same has been done for the synthetic telluric lines  to identify the line centers in wavelength space.
Synthetic spectra, by default, are constructed with Voigt profiles, and thus the 
Voigt profile (approximated as a Pseudo-Voigt profile) is the best choice for fitting. 
Observed telluric lines are also best represented by Voigt profiles. 
To avoid possible weak line contamination near the line wings, 
however, we decided to fit the wavelength range spanning from the line centroid to the line full width at half maximum (FWHM) depth. 
Many tests showed that at that level the lines are best fit by a Lorentzian profile,
and thus we selected the Lorentzian model  for telluric fitting instead of the Pseudo-Voigt model.

Finally, the wavelength solution is obtained by constructing a  third order polynomial $\chi^2$ fit between the 
precise CO$_2$ pixel coordinates of observed telluric lines and their wavelength centers from the synthetic spectra. 
The wavelength solutions obtained in this way are more precise than those 
from the CRIRES pipeline and can be used for obtaining precise RVs. 
Our RV precision, however, will always be limited by the individual 
telluric lines precision from the HITRAN catalog (5$-$50~m\,s$^{-1}$) as well as
the telluric variability. The HITRAN catalog precision can in principle be overcome using many lines so that the random errors average out, but the
telluric variability is a systematic error that affects all lines in the same way.
We tried to limit the random part  using as many lines as possible,
but blends between atmospheric and stellar lines sometimes did not leave many lines to work with.

\subsection{Spectra interpolation and telluric removal}
\label{Spectra interpolation and telluric subtraction}

Each wavelength solution for each chip and nodding position was interpolated
and resampled onto 10$^5$ regularly spaced pixels. 
Based on this oversampled solution, we interpolated the observed and synthetic spectra from their original, 
irregularly-spaced wavelength solutions onto the regularly spaced pixel grid. 
The high resolution LBLRTM telluric spectra was interpolated with a sampling factor $L$ $\sim$ three times higher than the original, 
while the observed spectra have $L$ $\sim$ 100 times the number of the original 1024 pixels.

Having both the observed and synthetic telluric spectra on the same wavelength scale is a great advantage.
In this way, the telluric lines in both spectra match in wavelength space, which allows us
to remove their contribution from each CRIRES spectrum. Synthetic spectra, however, have deeper and narrower 
lines, and must therefore be smoothed with the CRIRES instrument profile (IP) to match the observed 
line widths and depths. Instead of measuring each telluric line to identify the IP, we applied a more general strategy. 
By using a least squares algorithm, we compared the observed spectra, on the one hand, and 
the synthetic spectra convolved with a Gaussian kernel, on the other hand.
An initial guess of the width of the Gaussian kernel is estimated
from the spectrograph resolution ($\sim$~0.016~nm at $\lambda$~=~1594.5~nm); this is the only free parameter in the fit.
With this technique, the IP is obtained quickly and independently for each nod and detector. 
In our test we assumed that the IP is a simple Gaussian, although in reality 
the IP is a more complicated sum of several Gaussian profiles
coming from different optical parts of the spectrograph \citep{Butler, Bean2010}.

\begin{figure*}[htp!]
\begin{center}$
\begin{array}{cc}
\includegraphics[width=3.5in]{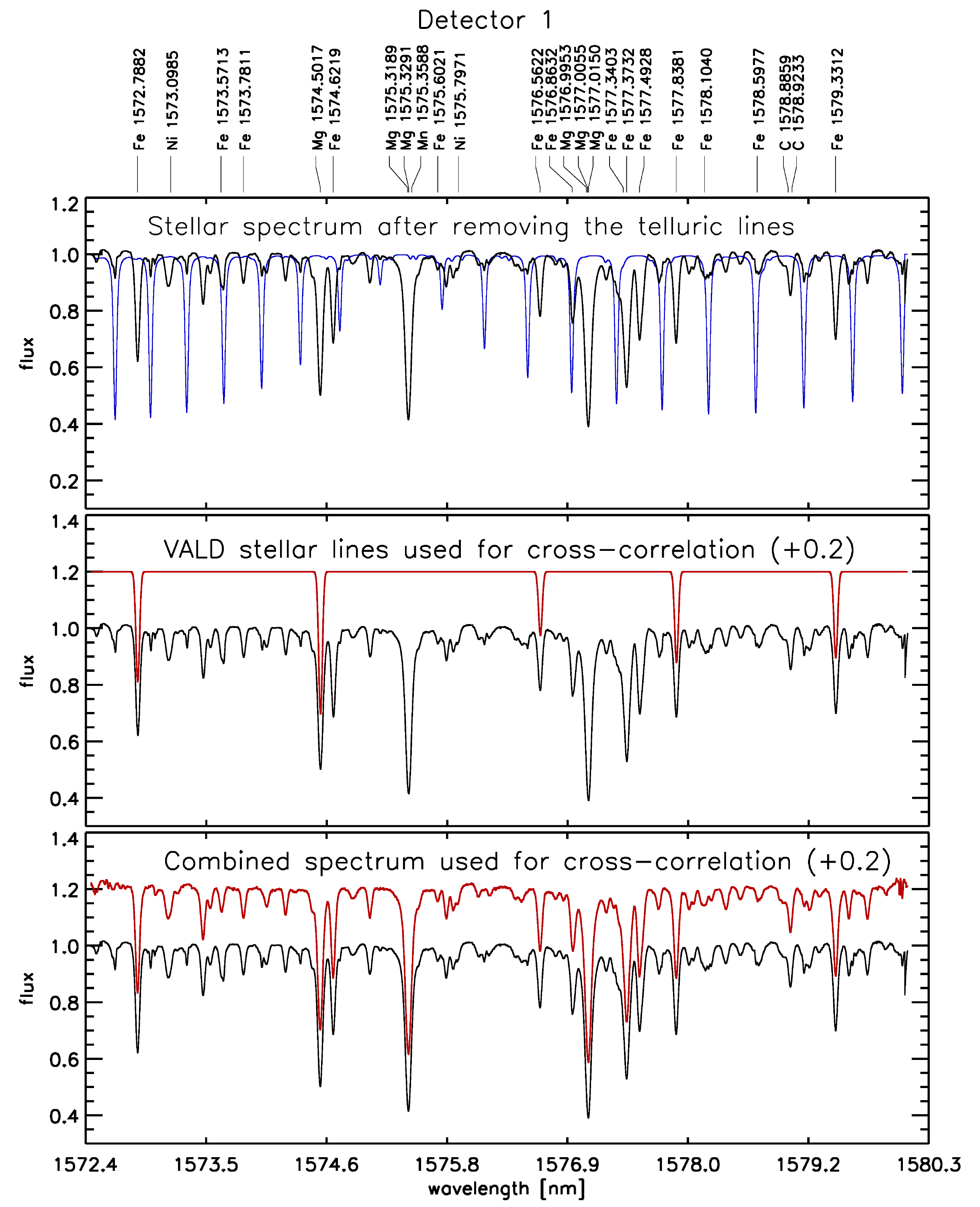} &
\includegraphics[width=3.5in]{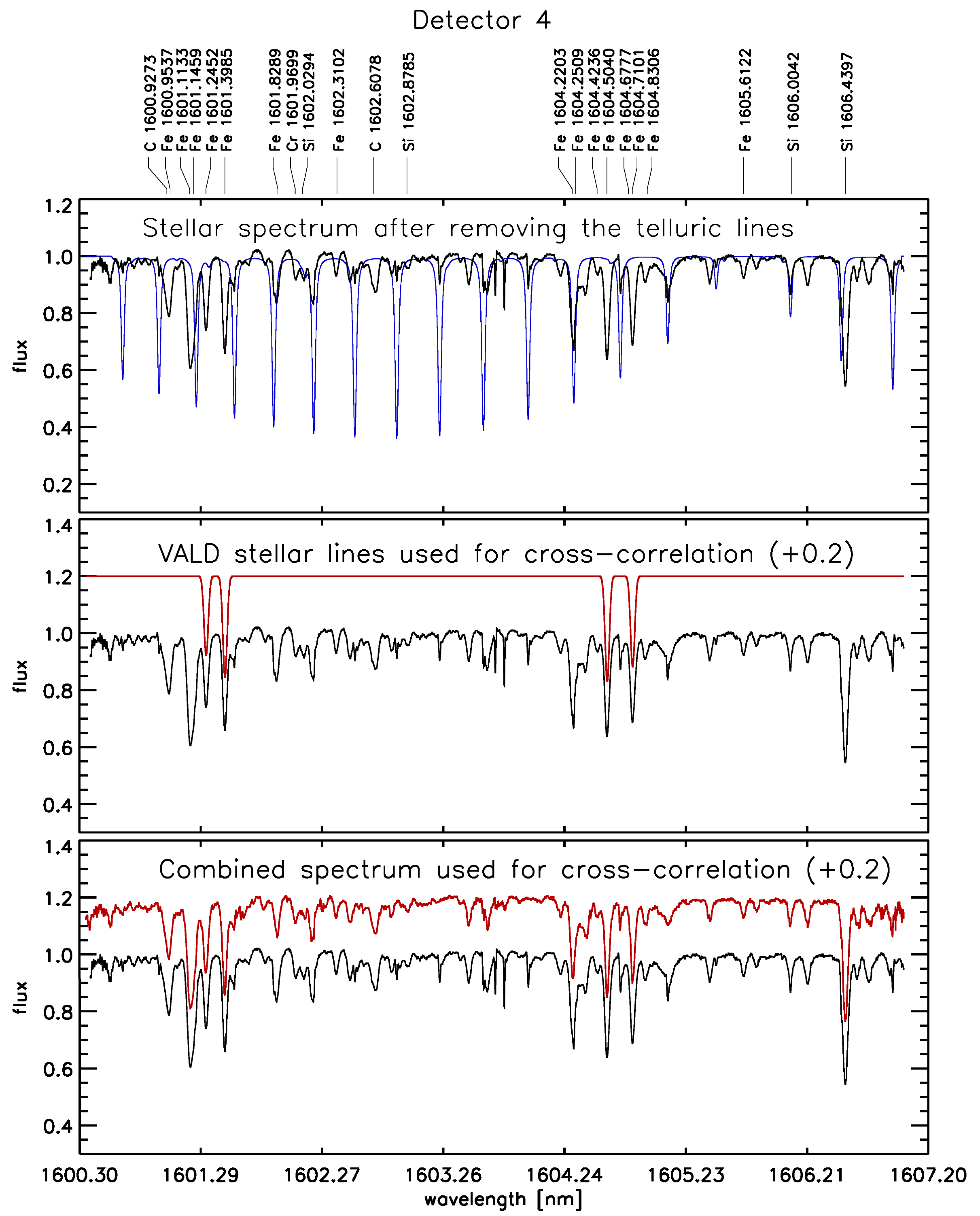}\\
\end{array}$
 
\end{center}
\caption{Top: spectra for detector one (left) and four (right) are divided by the synthetic telluric spectra (blue), removing the 
atmospheric contribution, and thus only leaving  the stellar lines (black). 
Many of these spectra were automatically identified in the VALD line catalog.
Middle: only the nonblended lines with well-defined profiles are used for obtaining the radial velocities 
via cross-correlation with a synthetic stellar template modeled from the VALD line catalog. 
Bottom: the telluric-free spectra from all observational epochs are later shifted and median-combined 
in one very high S/N stellar template, which is then used for cross-correlation.
} 
 
\label{FigGam:chip1-4c-cascade}  
\end{figure*}

Telluric lines may be removed in our pipeline by dividing the observed spectra 
by the synthetic telluric template convolved with the IP, leaving only the stellar lines.
The automated line algorithm was then once again applied to the remaining stellar lines.
The laboratory wavelengths of the identified stellar spectral 
lines were taken from the Vienna Atomic Line Database \citep[VALD,][]{Kupka} and obtained based on 
the target's $T_{\mathrm{eff}}$ and $\log g$. 
Fig.~\ref{FigGam:chip1-4c-cascade} (top panel) illustrates the final output from the spectral division. 
The figure shows many blended and unblended stellar lines remaining in the spectrum, 
but there are also many weak stellar lines for which we do not have any information from VALD.

We believe that dividing the spectra in this way provides us with the best
stellar template that can be extracted from the CRIRES spectra.
This method gives much better results than simply masking the telluric lines from the spectrum. 
With this approach, we remove the total atmospheric contribution from the science spectra.

\section{Obtaining the radial velocities}
\label{Obtaining the radial velocities}

To derive precise radial velocities, we adopted a method that combines steps from the 
least squares modeling as in the iodine method \citep{Butler} and the cross-correlation method \citep{Baranne}. 
Indeed, our wavelength calibrators are atmospheric CO$_2$
lines, which are always superimposed on the stellar spectra similar to the gas cell method. 
As we described in Section~\ref{Calibration and data reduction}, we were able to model the 
telluric lines and eliminate their contribution from the science spectra.
The ideal case would be to use the iodine method to model  the 
telluric and the stellar spectrum simultaneously, where one of the free parameters would be the Doppler shift.
In this kind of approach, the blends between telluric and stellar lines would be less problematic.
This approach, however, was initially not possible, because we did not have a proper 
stellar template spectrum in the near-IR region studied. 
Also, a simultaneous fit would be more complicated in our case because the telluric lines are variable
in contrast to the iodine lines, which are stable. 
Therefore, despite the fact that the four CRIRES detectors cover relatively small spectral regions, our
choice was to use a more simplified approach and we obtain radial velocities by cross-correlating 
the calibrated spectra with a proper stellar mask.

Initially, when we only had a few CRIRES observations for our targets, 
we obtained radial velocities by cross-correlating the stellar spectra with
a weighted binary mask \citep{Pepe}. Our mask was consistent with noise-free continuum level values
at those wavelengths where we did not identify stellar lines and at a noncontinuum
level for the stellar wavelength positions.
The mask also had adjustable aperture widths to take  the stellar line's FWHM  
into account or to select a pre-defined width, thus making the apertures box-shaped.
Even though we achieved relatively good results, we realized that a cross-correlation with weighted 
binary masks for only a few spectral lines available on each detector might not be optimal. 
Some near-IR RVs deviated considerably from the Keplerian model prediction, 
so it became clear that we might be far from the precision goal of 10$-$25 m\,s$^{-1}$.
The main problem was that for some epochs one or even both detectors had only two unblended 
stellar lines (minimum for our pipeline), which could be used to construct a cross-correlation function (CCF). 
If one of the two lines is slightly deformed by a nearby line, a delta function 
or box-like aperture mask  also leads to a biased CCF, and consequently 
to spurious velocity shifts. More lines or an accurate stellar template would mitigate
those problems. 

The line position precision, as calculated by state-of-the-art stellar atmosphere models such as 
PHOENIX\footnote{http://www.hs.uni-hamburg.de/EN/For/ThA/phoenix/index.html},
is usually worse than that obtained from the high S/N near-IR spectra themselves \citep{Figueira}, so
we decided to construct an accurate stellar template from our CRIRES observations.
The two different ways of constructing the stellar masks are explained below.

\subsection{CCF with noise-free stellar mask}
\label{CCF with noise-free stellar mask}

To construct a noise-free stellar mask from our observations we chose only single deep and sharp stellar lines
(with relative depths below 0.85 of the continuum level, as explained in Section~\ref{Wavelength calibration})
and modeled these using a Gaussian profile.
In fact, initially we did the modeling with Lorentzian and Pseudo-Voigt profiles,
but we found that the better fit and lack of complexity of the Gaussian profile suited us well. From this line modeling, we obtained the line FWHM and their individual spectral depths. 
In the next step, we shifted the identified reference stellar wavelengths by the target's 
absolute radial velocity\footnote{Either taken from SIMBAD 
(http://simbad.u-strasbg.fr/simbad/), or estimated with our pipeline.} and
Earth's barycentric motion \citep[accuracy better than~$\sim$~1~m\,s$^{-1}$,][]{roytman13}. 
With this approach, we took all the necessary line shifts into account, so that the synthetic stellar 
spectrum resulted in line positions very similar to those on the observed stellar spectrum.
We thus constructed a synthetic noise-free stellar spectrum in the same oversampled wavelength space 
as the observed stellar spectrum.

On Fig.~\ref{FigGam:chip1-4c-cascade} (middle panel) are shown the final stellar masks for detectors one and four. 
To save computing time, the cross-correlation is calculated for a range of about $\pm$ 5 km\,s$^{-1}$ 
(500 oversampled pixels) around the stellar line centers. 
The maximum of the CCF can be obtained down to subpixel level using $\chi^2$ minimization;
we tried fitting with a Gaussian, a Lorentzian, or a low-order polynomial, which fits a parabola
in the area of the maximum of the CCF.
Gaussian and Lorentzian models fit the CCF very well, but in contrast to the polynomial model 
they perform poorly around the CCF peak. 
The most likely reason for this is that these two models were applied over a wider region 
of the CCF when compared to the polynomial, which onl fits  around the maximum \citep{Prieto2007}.
Since we were interested in precisely modeling  the peak of the CCF, 
we selected the polynomial model  for deriving precise velocities.

This method of radial velocity computation was applied to the coadded spectra 
and to the spectra for nodding positions A and B at detectors one and four separately,
resulting in six different radial velocities for a given epoch. 
The final radial velocity was obtained by calculating a weighted mean of all radial velocities, 
where the weight was given by the median S/N of the extracted spectrum, which was taken from the FITS header. 
Using the S/N for weighting corresponds to the approach used by \citet{Butler}
for the combination of Doppler information from a large number of short spectral chunks.
Usually, the highest S/N ratio was achieved for the coadded spectrum
from detector one, and the lowest S/N belonged to one of the nodding positions at detector four.

\subsection{CCF with median combined stellar mask}
\label{CCF with median combined stellar mask}

As can be seen from Fig.~\ref{FigGam:chip1-4c-cascade}, there are many weak 
stellar lines on detectors one and four. On top of that there are a few very deep stellar lines
that according to the VALD line list are actually spectral doublets. 
These lines are difficult to model and cannot be used for precise
RVs in the context defined in Section~\ref{CCF with noise-free stellar mask}.

\begin{table*}[htp]
\centering

\caption{Comparison of best optical ($K_{\mathrm{opt}}$) and 
IR ($K_{\mathrm{IR}}$) RV semiamplitudes and companion between the optical and IR 
data r.m.s. values from these models. The sixth column
gives the proportion between the two semiamplitudes.}
\label{table:comp}

%\resizebox{0.99\textheight}{!}{\begin{minipage}{\textwidth}

 \begin{tabular}{lcccccl}
\hline
\noalign{\vskip 0.5mm}
\hline
\noalign{\vskip 0.9mm}

HIP  & $K_{\mathrm{opt}}$& $r.m.s._{\mathrm{opt}}$  & $K_{\mathrm{IR}}$& $r.m.s._{\mathrm{IR}}$ & $\kappa=K_{\mathrm{IR}}/K_{\mathrm{opt}}$& Reference \\  
     &      [m\,s$^{-1}$]&  [m\,s$^{-1}$]&     [m\,s$^{-1}$]& [m\,s$^{-1}$] &  &  \\  

\hline
\noalign{\vskip 0.9mm}

5364 b   & ~~51.1 $\pm$ 2.5 &  16.3 &~~65.7 $\pm$ 14.6 & 23.3 &1.29 $\pm$ 0.29 &\citet{Trifonov2014}\\
5364 c   & ~~52.9 $\pm$ 2.6 &  16.3 &~~71.5 $\pm$ 17.9 & 26.7 &1.35 $\pm$ 0.35 &\citet{Trifonov2014}\\
20889 b  & ~~93.2 $\pm$ 2.1 &  10.7 &~~89.5 $\pm$ ~~8.3 & 44.1 &0.96 $\pm$ 0.09&\citet{Sato2007} \\
31592 b  & ~~45.2 $\pm$ 4.7 & ~~7.3 &~~47.0 $\pm$ 10.8 & 31.8 &1.04 $\pm$ 0.24 &\citet{Wittenmyer2011}\\
34693 b  &  350.5 $\pm$ 3.4 &  21.1 &326.2 $\pm$ 13.6 & 39.9 &0.93 $\pm$ 0.04 &\citet{Mitchell2013}\\
37826 b  & ~~46.0 $\pm$ 1.6 &  10.9 &~~70.1 $\pm$ 12.6 & 15.4 &1.52 $\pm$ 0.27 &\citet{Reffert}\\
60202 b  &  296.7 $\pm$ 5.6 &  30.0 &277.8 $\pm$ 26.1 & 55.9 &0.94 $\pm$ 0.09 &\citet{Liu2008}\\
88048 b  &  288.4 $\pm$ 1.2 & ~~9.0 &253.4 $\pm$ 23.1 & 20.9 &0.88 $\pm$ 0.08 &\citet{Quirrenbach}\\
88048 c  &  175.2 $\pm$ 1.4 & ~~9.0 &\ldots           &\ldots&\ldots          & \citet{Quirrenbach}\\
114855 b & ~~91.0 $\pm$ 2.3 &  18.9 &120.7 $\pm$ 24.3 & 39.8 &1.33 $\pm$ 0.27 &\citet{Mitchell2013}\\
\hline
\noalign{\vskip 0.5mm}
%\hline
%\noalign{\vskip 0.5mm}

\end{tabular}
%\end{adjustwidth}
%\end{minipage}}
\end{table*}

After the telluric removal on all available spectra, we median combine all frames 
into another stellar template to be used for cross-correlation. This method has several major advantages:

\begin{enumerate}

   \item  Median combination  removes the telluric artifacts still present in the individual frames 
   after the division by the theoretical telluric spectra.
 
   \item All available spectral lines can be used, even unknown lines and unresolved doublet lines. 
   This is not possible by cross-correlating with the noise-free template.
 
   \item Cross-correlation between two identical functions  leads to a significant CCF 
   maximum, so that one can very precisely obtain the Doppler shift.  
\end{enumerate}

This method, however, requires several challenging steps. 
Individual spectra have different wavelength shifts and sampling, and thus need to be shifted carefully
before median combination.
Resampling is done by interpolating each spectrum to the most precise wavelength solution
achieved so far (i.e.,\ most telluric references have been used). 
Each spectrum is then shifted exactly by the difference to the reference spectrum and, finally, a median combination is applied.
The result is a high S/N stellar mask with the most precise wavelength solution for a given target.

This method was only applicable after we had acquired 5$-$8 epochs for each target. 
The final median combined stellar template spectrum for each target 
on detector one and four is shown in Fig.~\ref{FigGam:Mask_Median1} and Fig.~\ref{FigGam:Mask_Median4}.
The RVs seem to be more precise from detector one than from detector four,
where the lower S/N and the abundant and faint stellar lines influence the CCF.

The resulting radial velocity for each epoch is obtained with the same cross-correlation steps and weighted
combination of individual frame RVs as explained in Section~\ref{CCF with noise-free stellar mask}
for the noise-free stellar template. 
The measured near-infrared radial velocities for all stars using this method are listed in Table \ref{table:RVs}.

\subsection{Error estimation}
\label{Errorestimation}

The RV uncertainties depend on the number of telluric lines used for the calibration 
(i.e.,\ the reproducibility of the wavelength solution), the lines used for cross-correlation,
and the stellar flux from each individual star.
We did not assess these systematic errors quantitatively nor did we consider the HITRAN and VALD line errors individually.
Instead, for each observation we used one combined error, which is based on the RV dispersion
from all nodding and coadded positions.
The RV error computed in this way varies greatly from epoch to epoch, with a mean value of about $\sim$ 25 m\,s$^{-1}$. 
The individual error for each observation is listed in Table \ref{table:RVs}.

Based on our results, we estimated the total error of our CRIRES measurements to be not more than 40 m\,s$^{-1}$.
This estimation was further confirmed by the total r.m.s.\ dispersion of the near-IR velocities around the 
best-fit model obtained for each target from the Lick velocities.

\section{IR radial velocity analysis}
\label{IR data analysis}

\subsection{Consistency between optical and IR data}
\label{Comparison with the optical}

Eight stars in our sample have published planetary companions, including two two-planet systems. For those ten 
planets\footnote{For simplicity we also use the term planet for deuterium burning mass objects (brown dwarf)
as defined in \citet{Reffert2014}.} the five spectroscopic orbital parameters characterizing a Keplerian orbit are known rather accurately.  
To test whether the CRIRES radial velocities support the planet hypothesis, we fitted Keplerian
orbits to the CRIRES radial velocities for each star as well. However, since we have far fewer data points from CRIRES
than in the optical,   sometimes with incomplete phase coverage, we only fit for the RV semiamplitude 
as well as the RV zero point;  we also keep the other four spectroscopic parameters (period, periastron time,
eccentricity, and longitude of periastron) fixed at the values derived from the optical radial velocities. 
We note that our observations with CRIRES were scheduled to sample the RV signal as well as possible in phase over the orbital period. 
With this approach, we tried to avoid uneven sampling of the data because it can lead to possible zero velocity crossings
and a poorly resolved periastron passage \citep[e.g.,][]{Cumming2004}, which can lead to highly uncertain amplitude ratios.

For the two stars with two planets, the fitting was done one planet
at a time, while the RV signal of the other planet as determined from the optical data was 
subtracted. This approach assumes that the RV signal of the subtracted planet is consistent in the
optical and the IR; when this is not the case then this inconsistency would show up in the comparison 
of the RV signals for the other planet in the system.

As in the optical, we quadratically added an RV jitter term to the RV measurement errors 
to account for stellar noise (pulsations). We did not make any attempt to derive a different value for the jitter
term from the CRIRES data, but used the jitter term derived from the optical
data. For the eight stars investigated here, the RV jitter term varies
from about a third of the value of the IR RV measurement error up to about twice its value.

\begin{figure}
\includegraphics[width=3.3in]{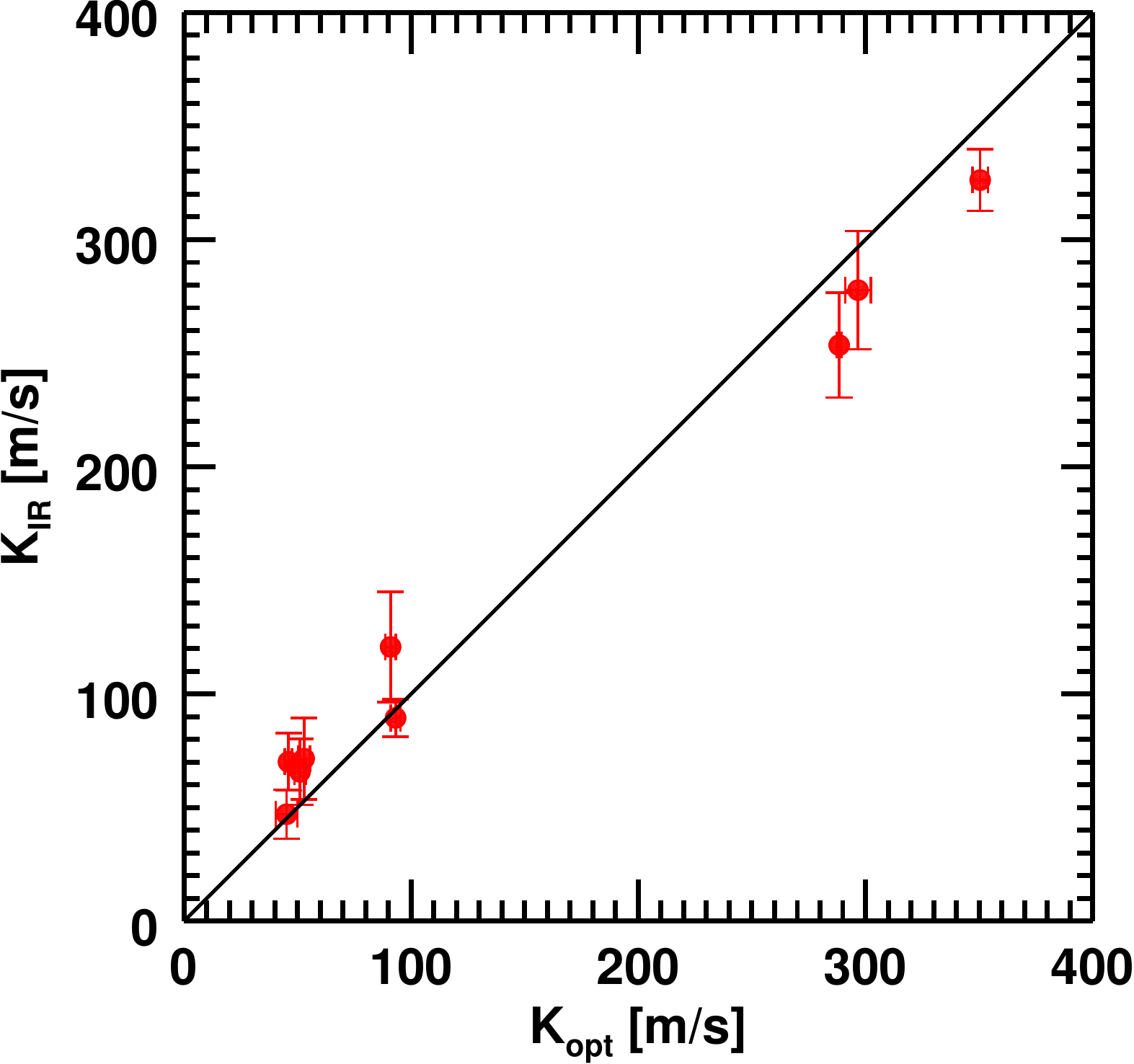} 
\caption{\label{fig:comp}Comparison of the optical ($K_{\mathrm{opt}}$) and infrared (K$_{\mathrm{IR}}$) 
RV semiamplitudes. The straight line corresponds to $K_{\mathrm{opt}} = K_{\mathrm{IR}}$.
Overall we observe a good correspondence between $K_{\mathrm{opt}}$ and $K_{\mathrm{IR}}$;
the largest deviations are at the 1.5$\sigma$ level.}
\end{figure}

The resulting RV semiamplitudes in the optical ($K_{\mathrm{opt}}$) and in the IR ($K_{\mathrm{IR}}$) are
given in Table~\ref{table:comp} together with their errors derived from
$\chi^2$ fitting. Table~\ref{table:comp} also shows the r.m.s. values of the optical and IR 
data from the best optical and IR models, respectively.
The last column in the table gives the proportion $\kappa$ 
between those two RV amplitudes: $\kappa = K_{\mathrm{IR}}/K_{\mathrm{opt}}$, 
along with its error derived following simple error propagation.
The RV semiamplitudes $K_{\mathrm{opt}}$ and $K_{\mathrm{IR}}$ are also
shown in Fig.~\ref{fig:comp}. The solid line corresponds to $\kappa=1$, i.e.,\
the same RV semiamplitude in both wavelength regimes. 
In Fig.~\ref{criresvels}, the resulting Keplerian radial velocity curves for the
optical and the IR wavelength regime are shown together with the CRIRES measurements. 

The RV semiamplitude of the outer planet in the HIP~88048 system is not constrained 
by the CRIRES data, which leaves nine planets in eight systems for our analysis.
As one can see, the optical and the IR data are consistent (i.e.,
$\kappa\,\approx\,1$) at the 1--1.5$\sigma$ level for all nine investigated planets. 

In the case of a radial velocity signal, which is at least partly due to stellar
noise (starspots, pulsations, etc.), one might expect different RV amplitudes or
phases in the optical and in the infrared since the contrast between the 
various regions on the stellar surface is wavelength dependent.
The agreement between optical and infrared amplitude further supports the 
planetary hypothesis in all eight systems, as expected further tests might be required to
undoubtedly confirm the planet interpretation. 
We will discuss the other systems individually in forthcoming papers.

\subsection{IR data phase coverage}
\label{phase coverage}

Given the relatively sparse IR data sampling for our targets 
 as well as the moderately large IR RV errors, we investigated
whether this might lead to systematic errors in the fitted RV semiamplitudes.
We simulated 10\,000 alternative IR data sets for each of the eight stars.
We assumed the optical orbit to be the correct data set and we randomly 
simulated IR data drawn from a normal distribution around the best
optical fit, but 
at the same observational times and with errors as in our original CRIRES measurements.
For each simulated data set, we fitted the RV semiamplitude as described in
Section~\ref{Comparison with the optical},
keeping the rest of the orbital parameters fixed at the optical solution.
The resulting distribution of RV semiamplitudes was fitted  
with a Gaussian, from which we obtained the mean RV semiamplitude and its standard deviation.
 This test should tell us whether the incomplete phase coverage coupled with the
relatively large IR RV errors would lead to any systematics in the recovered RV semiamplitudes.

We found that for all targets listed in Table~\ref{table:comp} the RV
semiamplitude could be recovered from the simulated data without any systematic offsets.
The biggest difference  we found between the optical and the simulated RV
semiamplitude was $\sim$ 0.3 m\,s$^{-1}$, which is completely insignificant.
We conclude that there are no systematic differences in $\kappa$ due
to the sparse sampling and large error of the IR data set.

\section{Discussion}
\label{Results and Disc}

\subsection{RV precision}

As discussed in Section~\ref{Errorestimation}, the RV precision achieved with CRIRES spectra
depends strongly on the number of telluric and stellar lines used for wavelength calibration and cross-correlation, respectively.
In this context we find that giants of spectral type G8\,III -- K2\,III are the 
best targets from our CRIRES survey in terms of radial velocity precision, when we cross-correlate with the noise-free stellar mask
(see \S\ref{CCF with noise-free stellar mask}). These stars have a small number of deep stellar lines, and hence, 
the problematic line contamination with CO$_2$ lines is minimized. We were able to use a maximum number of telluric lines for precise 
wavelength calibration and almost in any epoch we had enough unblended stellar lines to obtain radial velocities.
We achieved similar results for these stars when we cross-correlate the telluric free spectra with the median-combined stellar mask
 (see \S\ref{CCF with median combined stellar mask}). 
For G8\,III -- K2\,III giants we achieved an overall mean RV precision of about 25 m\,s$^{-1}$.
Also, it is worth mentioning that G8\,III -- K2\,III giants have in general lower 
levels of stellar RV jitter than later K giants, and the velocity precision achieved was adequate to test 
the near-IR velocities for agreement with optical data, and hence, to test for the presence of a planetary companion.

Although both methods worked well for G8\,III -- K2\,III giants, the noise-free mask failed to give reasonable 
velocities for K3\,III -- K4\,III giants. The reason for that is the abundance of stellar lines typical for late K stars
 (see Fig. \ref{FigGam:Mask_Median1} and Fig. \ref{FigGam:Mask_Median4}).
Heavy line blending mutually excluded large number of lines usable for the wavelength solution and cross-correlation.
We find that for many cases in given epochs the RV dispersion between the individual detector 
nods is too large to derive an adequate RV precision.
Unlike the noise-free stellar mask, the median mask is less sensitive to blended stellar lines 
(since we use all available stellar lines to construct the CCF) and generally gave better results for these stars. 
Although in some cases the low number of unblended telluric lines was still a problem 
for constructing a precise wavelength solution, the full set of stellar lines used 
in this method led to a significant CCF peak from which we could obtain the RVs with a decent precision.
 The only way to obtain reasonable RVs for K3\,III -- K4\,III giants was with the median stellar template. 
Using this method, we achieved mean precision of 20 m\,s$^{-1}$, which is even slightly better than that for G8\,III -- K2\,III.

\subsection{Substellar companions}

\begin{figure*}[htp!]
 \begin{center}$
\begin{array}{ccc}
  \includegraphics  [width=6cm]{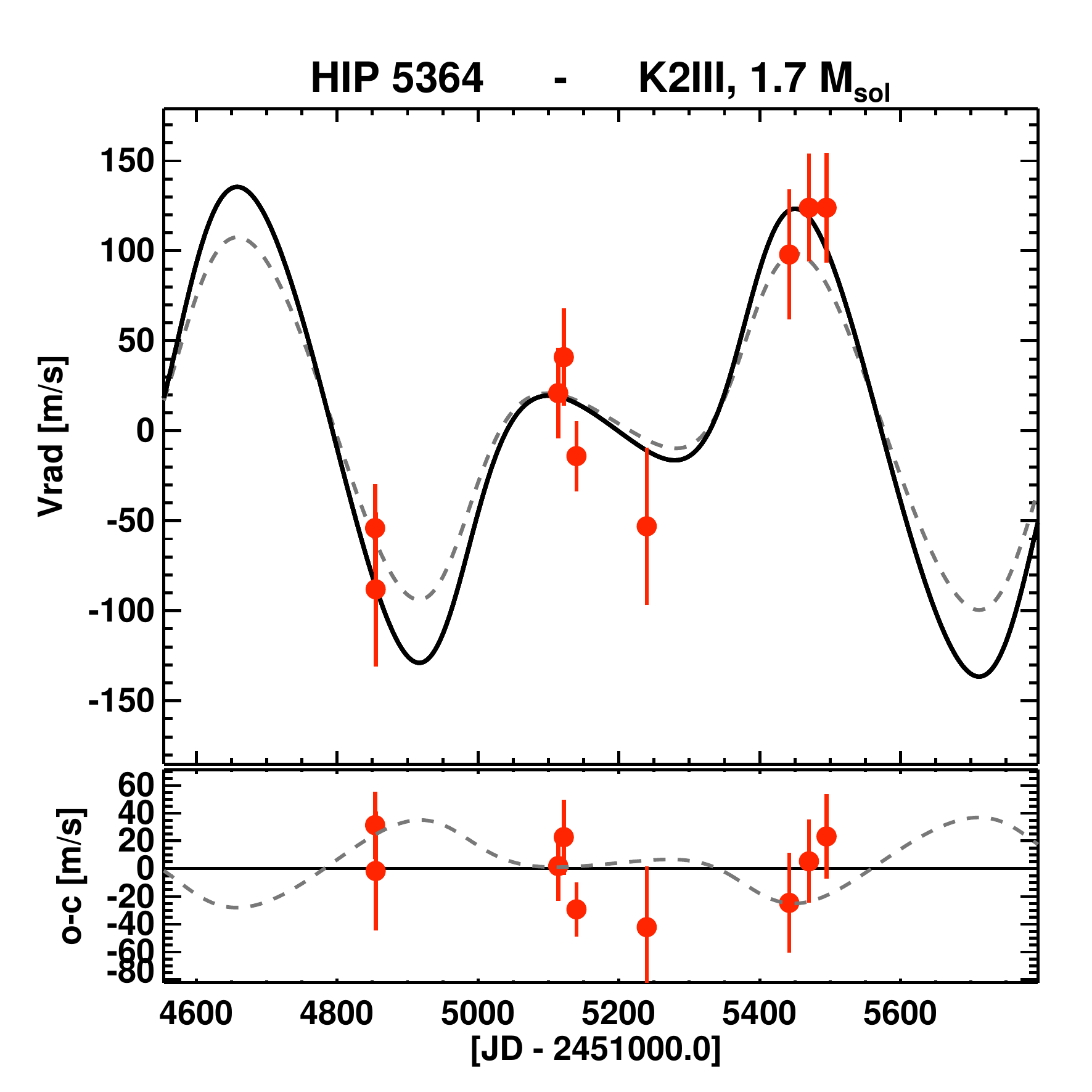} 
  \includegraphics  [width=6cm]{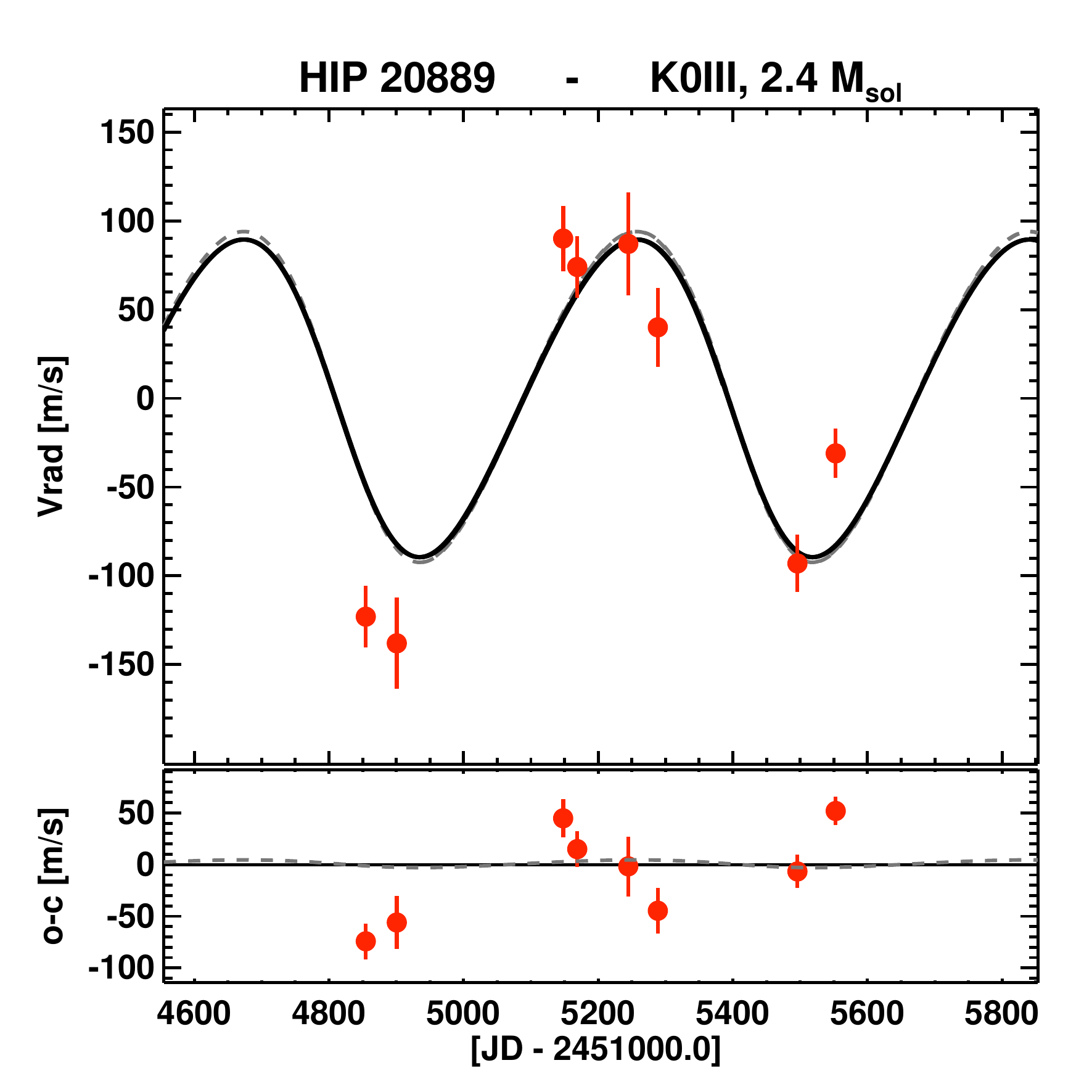} 
  \includegraphics  [width=6cm]{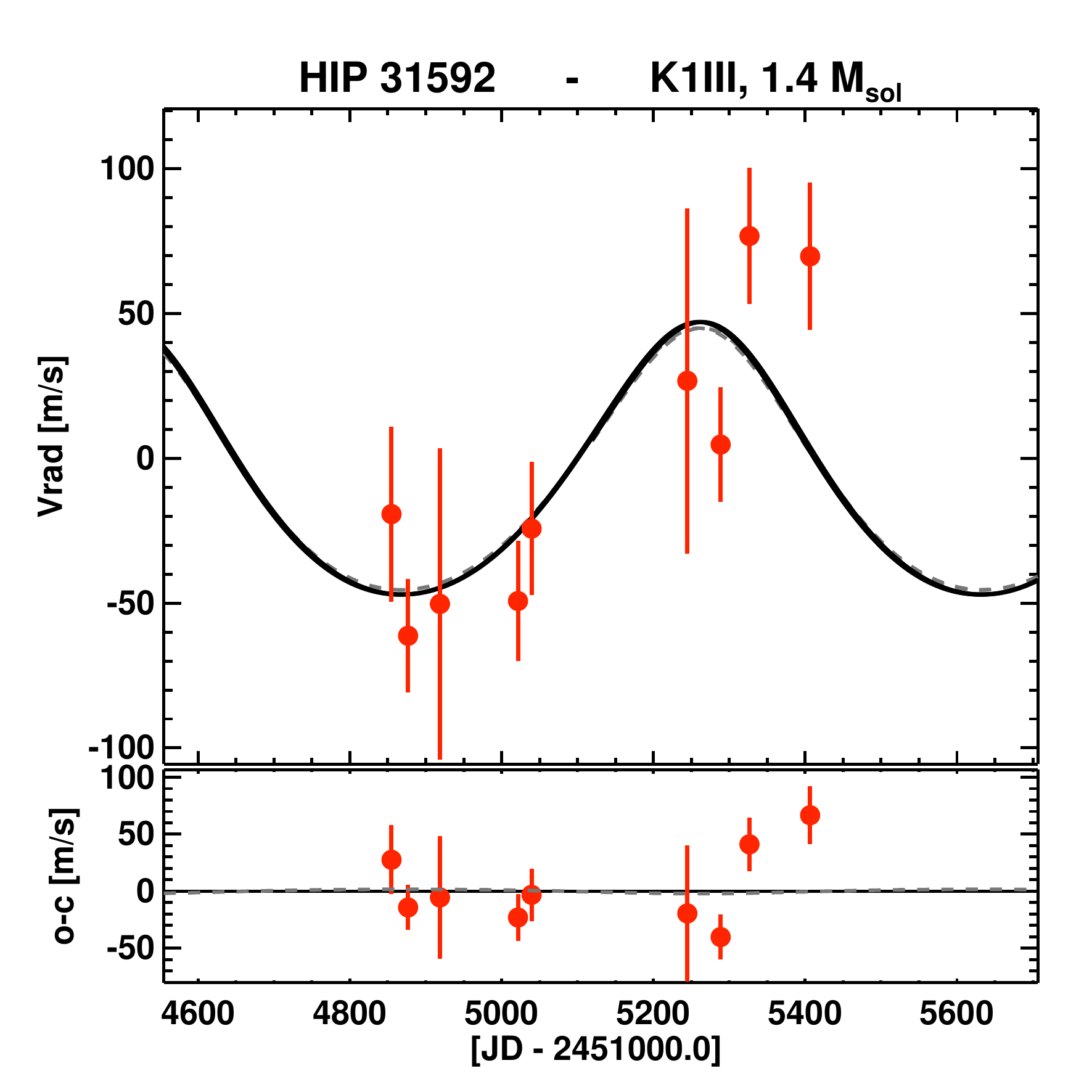} \\
  \includegraphics  [width=6cm]{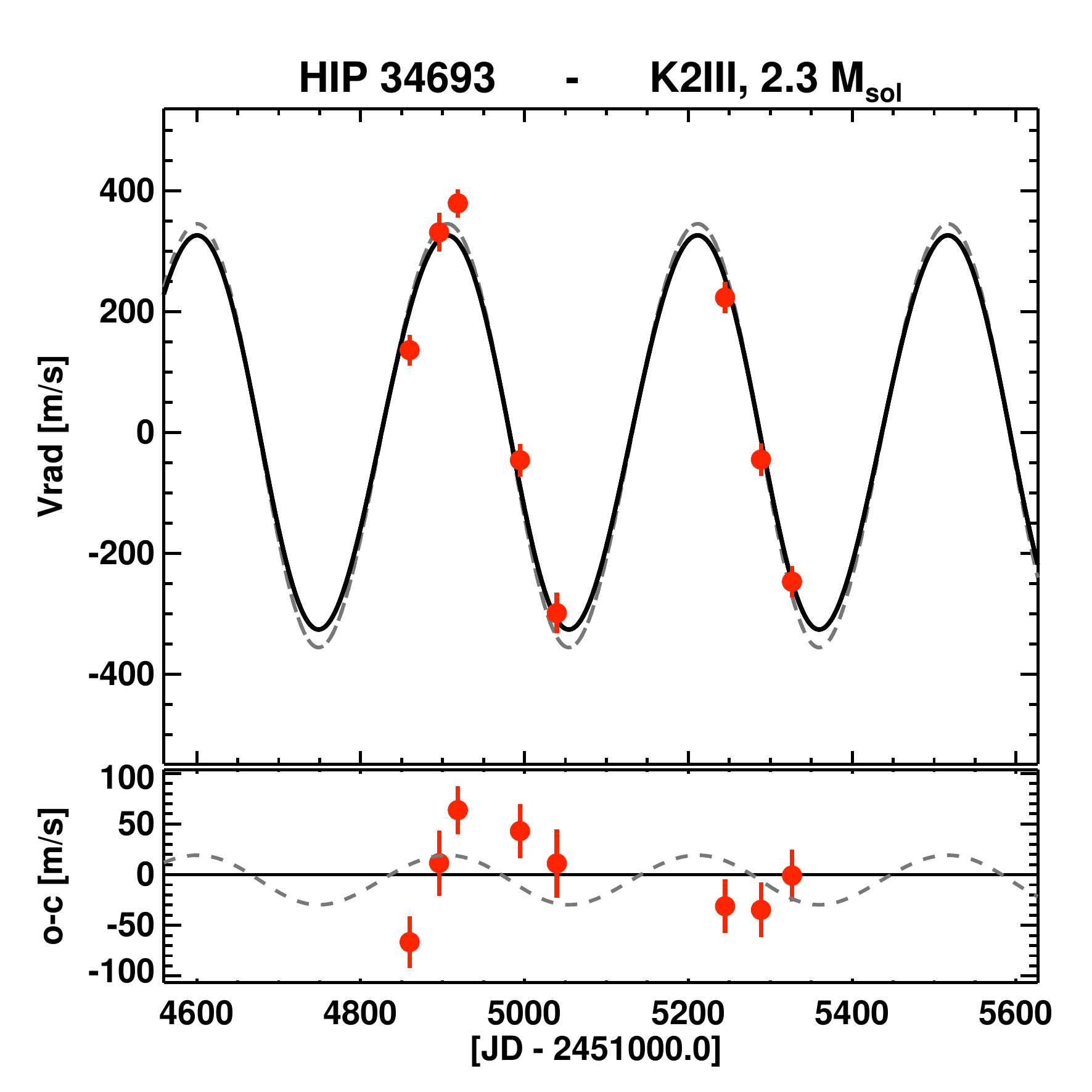} 
  \includegraphics  [width=6cm]{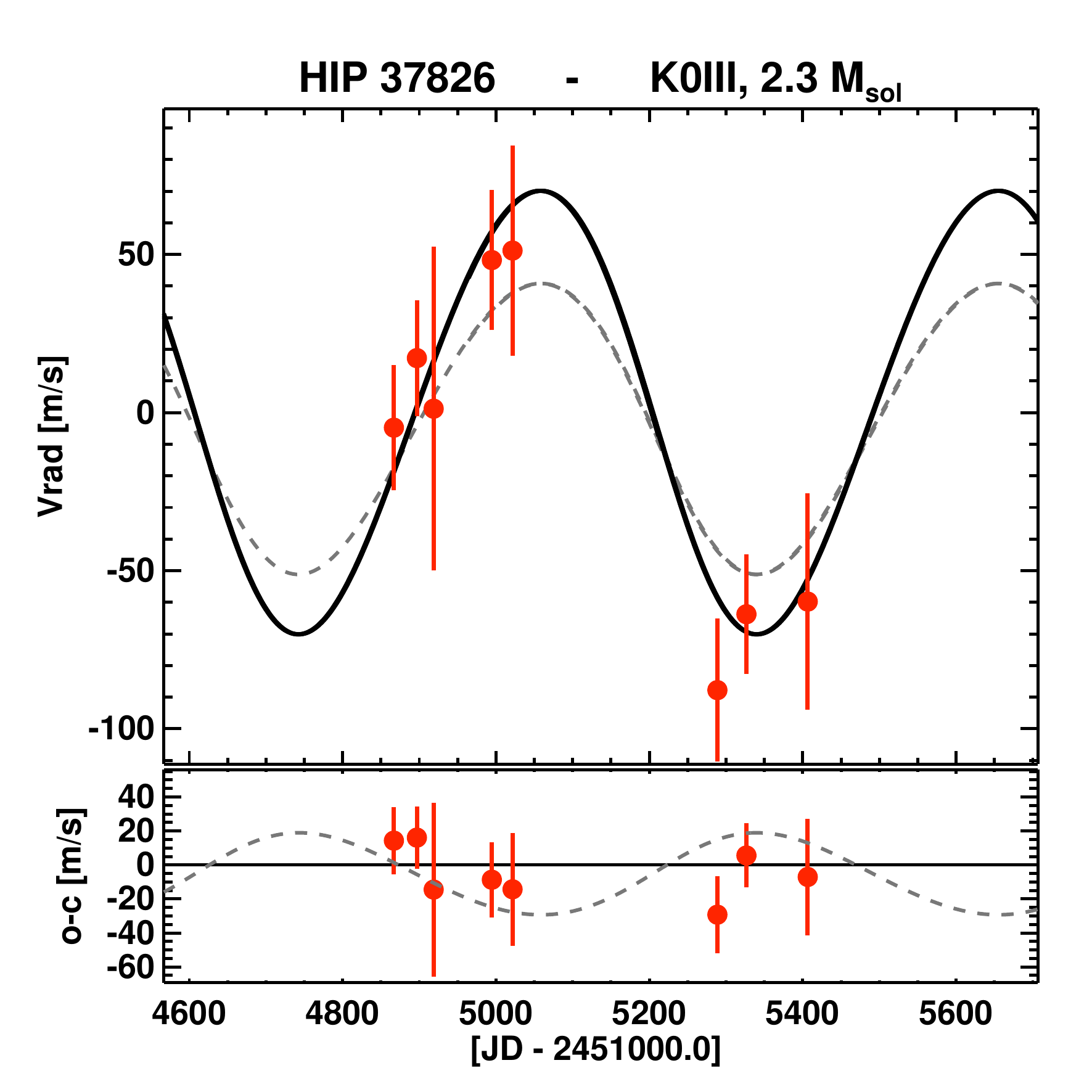} 
  \includegraphics  [width=6cm]{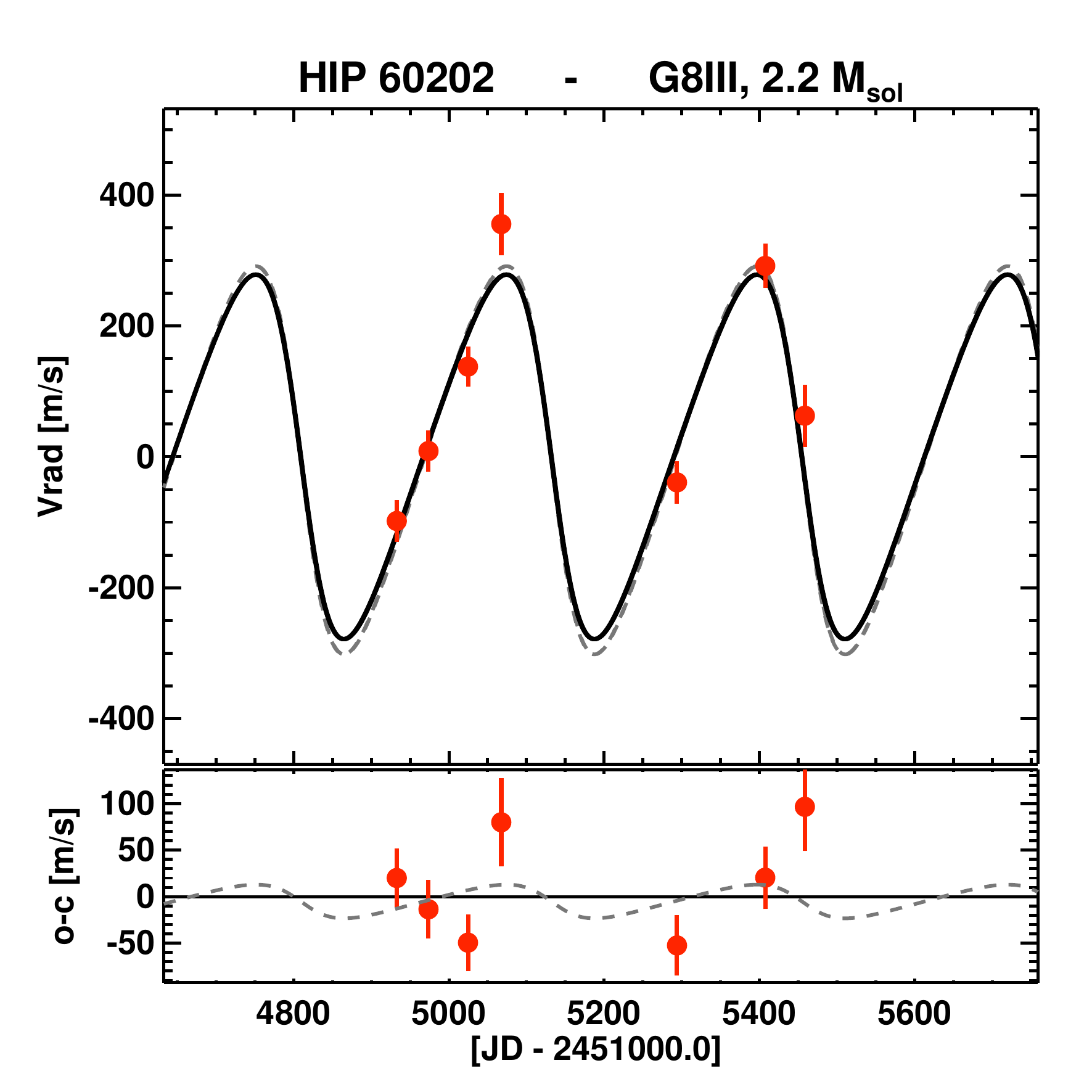} \\
  \includegraphics  [width=6cm]{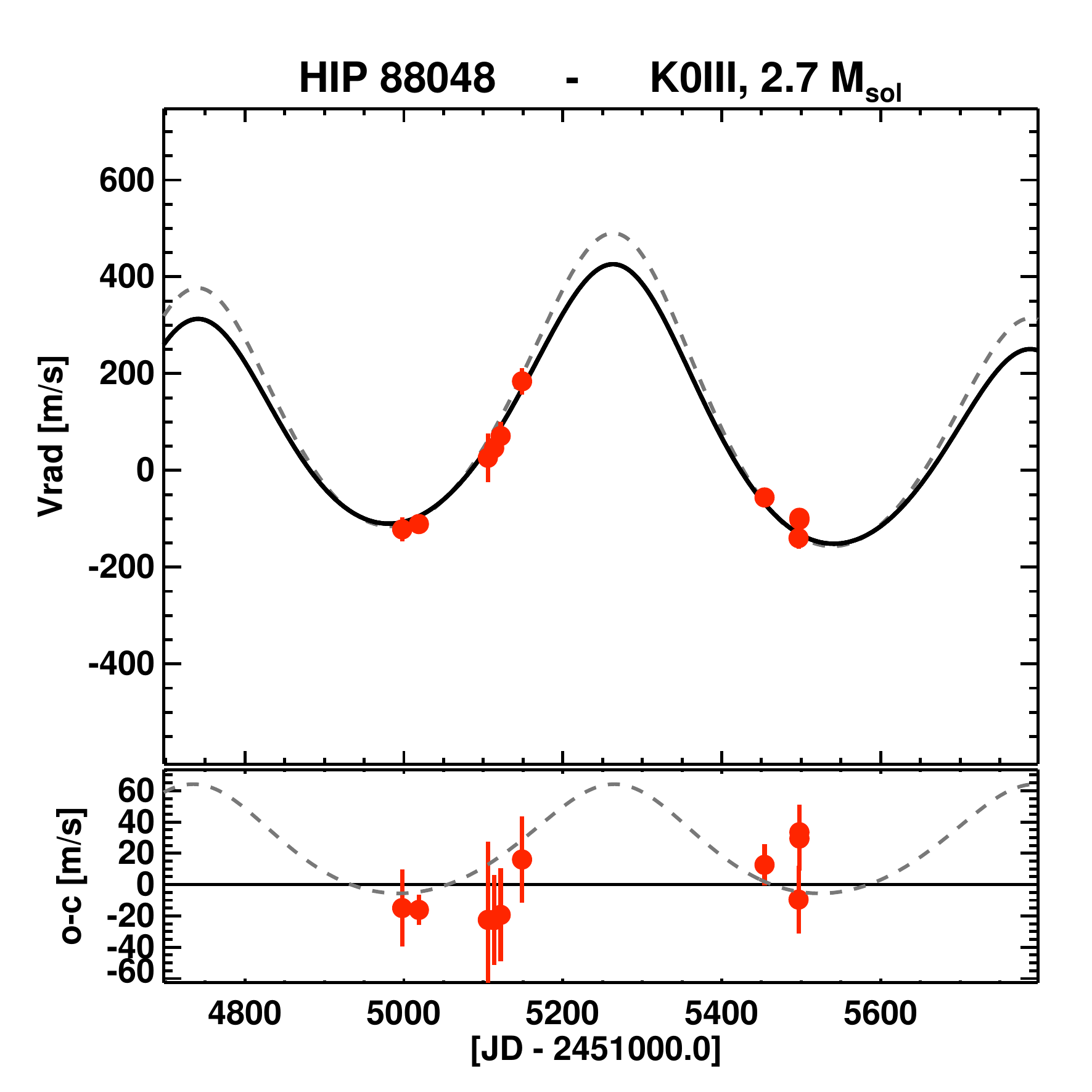} 
  \includegraphics  [width=6cm]{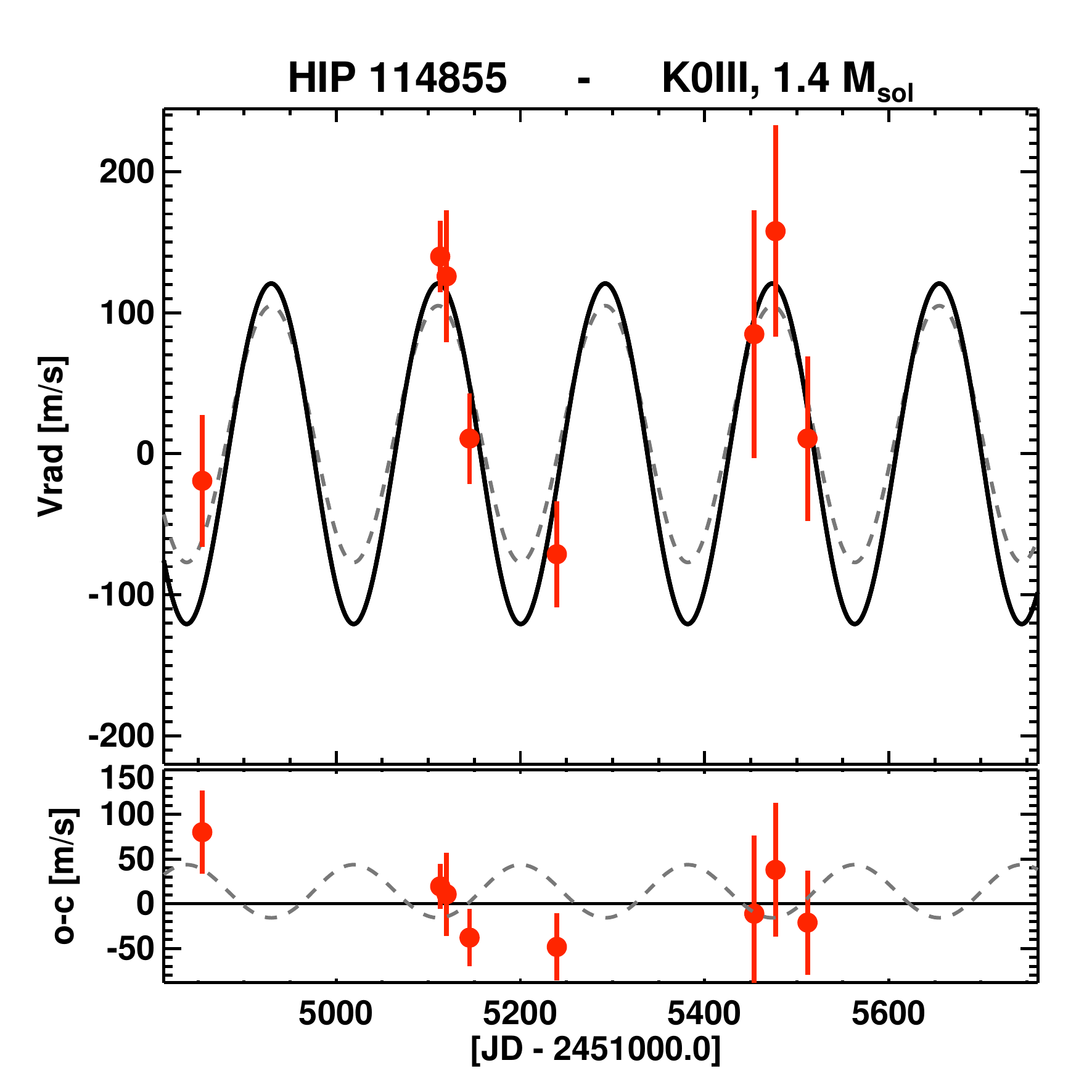}   \\

  \end{array}$
 \end{center}
 
  \caption{CRIRES near-IR radial velocities for the targets known to have planetary companions.
  Two best fits to the CRIRES data are overplotted:
  the model for the best-fit value of $\kappa$, where $K_{\mathrm{IR}}$ is a free parameter 
  is plotted with a solid black line, while the gray dashed line is for $\kappa$ = 1, which is the best
  Keplerian model obtained from the literature data (optical).
  Bottom~panels show the residuals around the best-fit value~of~$\kappa$. The dashed line in the residual 
  panel illustrates the difference between the two models.
  }
  \label{criresvels} 
  \end{figure*}

Lick and CRIRES data for HIP~114855 and HIP~34693 have already been published in \citet{Mitchell2013}, while 
the multiple planetary system around HIP~5364 was extensively discussed in \citet{Trifonov2014}. 
Here we present improved CRIRES velocities for these targets 
based on cross-correlation with the median-combined mask, superceding  earlier results. 
Results from our near-IR study based on the new CRIRES velocities 
for these stars are given in Table \ref{table:comp} and Fig. \ref{criresvels}.
All three stars have consistent optical and near-IR semiamplitudes at the 1$-$1.5$\sigma$ level. 
 The relatively sparse near-IR sample
for HIP~114855 and HIP~34693 has excellent phase coverage showing consistency with the best optical Keplerian model.
HIP~5364 has a rather good near-IR phase coverage with data mostly sampled around the two-planet signal
extrema, but more near-IR RVs for later epochs would be highly desirable. 
As was demonstrated in \citet{Trifonov2014}, Lick and CRIRES data are  
more consistent with a two-planet dynamical model rather than a simple double Keplerian model, and thus strongly argue
for the presence of gravitationally interacting planets. 
However, for longer time spans the difference between the Keplerian and the dynamical model is 
much larger, and thus more data  certainly helps to reveal HIP~5364 system's architecture in more detail.
We conclude that HIP~114855, HIP~34693, and HIP~5364 have secure planets.

HIP~60202 has a substellar companion with minimum mass in the brown dwarf regime 
($m_{\mathrm{b}}$ $\sin i \approx$ 17 $M_\mathrm{Jup}$), first reported in \citet{Liu2008}. 
More optical velocities from Lick for this star will be published in Mitchell et al. (in prep.). 
Our CRIRES velocities for this star are consistent
with the Keplerian model based on the combined optical velocities from Lick and those published in \citet{Liu2008}.
We found that compared to the optical model the near-IR data favor a slightly
smaller RV semiamplitude ($\kappa$ = 0.94 $\pm$ 0.09), but both data sets are clearly consistent with each other.
As can be seen in Fig.~\ref{criresvels}, the sparse near-IR data set does not cover the time of the predicted RV minimum, which is
most likely the reason for the lower~$\kappa$.
We conclude that the IR data support an orbiting companion around HIP~60202.

The optical Doppler velocities for the K0\,III giant HIP~88048 are clearly consistent with two massive companions on noncircular orbits.
Based on our Lick data, two brown dwarf companions were first discovered by \citet{Quirrenbach} and later confirmed by \citet{Sato2012}. 
The Keplerian nature of the optical velocity data was generally accepted, since
other phenomena intrinsic to the star are very unlikely to be responsible for the Doppler signal.
The period of the outer companion exceeds several times the time span of the CRIRES observations, and thus at this stage 
we cannot derive any $\kappa$ value for the second RV signal.
The near-IR data cover only about one full period of the inner planet and clearly follow the optical model, but with 
 poor phase coverage at the predicted RV maximum (see Fig.~\ref{criresvels}). 
Our semiamplitude analysis resulted in $\kappa$ = 0.88 $\pm$ 0.08, which is about 1.5$\sigma$ from the best optical fit.

We also confirm the presence of planetary companions around HIP~20889 \citep{Sato2007} and HIP~37826 \citep{Reffert, Hatzes}.
For HIP~20889, the near-IR data cover one full period and the
velocity amplitude fully agrees with the amplitude from the optical resulting in $\kappa$ = 0.96 $\pm$ 0.09. 
For HIP~37826, the near-IR velocity amplitude is larger than the predicted amplitude with $\kappa$ = 1.52 $\pm$ 0.27, but
the near-IR value is still within 2$\sigma$ from the best optical semi-amplitude.
The reason for the larger $\kappa$ is most likely the relatively large near-IR RV errors compared to the optical semi-amplitude for HIP~37826.
Another complication might be the phase coverage; there are several measurements around the minimum and maximum RV, but in between there is a gap.
The CRIRES data follow the Keplerian predictions for these stars; the near-IR RV signal has slightly larger 
velocity dispersion than the optical data, as expected.

The CRIRES data are best suited for testing RV models that assume brown dwarf mass objects in orbit (HIP~34693, HIP~60202 and HIP~88048),
since these stars have large RV semiamplitudes and the CRIRES radial velocity uncertainties are still adequate to detect the periodic RV signal. 
It was more challenging for planetary mass objects, since their RV semiamplitude 
was usually only between two and four times of the achieved CRIRES measurement precision.
This was the case for HIP~31592, where the optical semiamplitude is on the order of $\sim$45 m\,s$^{-1}$ and it is on the same
order with some of the obtained errors from CRIRES (see Fig.~\ref{criresvels}).
Despite the large errors, however, the near-IR data for HIP~31592 are consistent and are best fitted with 
an amplitude that is about the same as in the optical ($\kappa$ = 1.04 $\pm$ 0.24).
These results imply that the planet announced in \citet{Wittenmyer2011} is very likely real,
although this star will benefit from more precise RV data with better phase coverage.

\section{Summary}
\label{Summary}

The main goal of our study was to confirm or disprove the planetary origin of radial-velocity variations observed in G and K giants. 
For this purpose, we compared a set of high precision optical and near-IR radial velocities and searched for 
consistency between the two wavelength domains.

For our test we selected 20 G and K giants that we extensively observed for more than a decade at Lick observatory
and  that all show periodicity in the optical RV data. For some of the stars orbiting planets have been
announced or will be published in the future, and for some there are
indications (from the optical data alone) that the RV periodicity might also be caused by stellar activity.
For all stars however the radial velocity signal is very persistent, so that we were able to
fit Keplerian orbits consistent with one or two orbiting substellar companions.

We obtained precise near-IR radial velocities with ESO's CRIRES spectrograph for these stars and in
this paper we presented our observational setup, data reduction strategy, and results. 
We selected a spectral region in the H-band, which is known to have abundant
atmospheric CO$_2$ lines that we used for wavelength calibration.
Only detectors one and four were used for our analysis. 

To achieve a precise wavelength solution, we adopted the CO$_2$ line centroids 
from a synthetic telluric spectrum constructed from a line-by-line radiative transfer code (LBLRTM), 
which uses as an input the HITRAN atmospheric line catalog and the ambient conditions at the time of the observations.
By performing a division between the synthetic atmospheric spectrum and the science spectrum, we precisely removed the telluric line
contamination, leaving only the stellar lines.

Finally, we derived the RVs by cross-correlating the observed spectra with noise-free stellar masks constructed from the 
VALD theoretical wavelengths and the modeled nonblended stellar line profiles from the observed spectra.
This method worked well for G8\,III -- K1\,III and some K2\,III giants, but failed to give reasonable velocities for later K giants. 
Therefore, once we had enough observational epochs for each target we combined all available target spectra into a master template. 
We cross-correlated this high S/N stellar template with each observed spectrum
 to extract the maximum Doppler information from all stellar lines in the spectra.
This improved strategy gave excellent results for almost all spectral types in our sample, including the late K giants.

The RV precision varies considerably from epoch to epoch and from star to star; the mean RV error is about 25~m\,s$^{-1}$.
The achieved RV precision was adequate to address our scientific question only for the stars with larger RV amplitudes.

The near-IR radial velocities are in general in excellent agreement with the optical
data; we did not find a single system where the two data sets are inconsistent with each other.
For a small number of stars the interpretation is unclear, since we would need more precise data to perform the comparison.
For all eight stars that we investigated in detail, the derived near-IR radial velocities agree well with the Lick data. 
They follow the best Keplerian model predictions and in particular they are consistent in amplitude,
so that we confirm the planets in the following systems (cf.~Table~\ref{table:comp}):
HIP~5364 ($\eta$~Cet), HIP~20889 ($\epsilon$~Tau), HIP~31592 (7~CMa), HIP~34693 ($\tau$~Gem), 
HIP~37826 ($\beta$~Gem), HIP~60202 (11~Com), HIP~88048 ($\nu$~Oph) and HIP~114855 (91~Aqr).
We conclude that the vast majority of the planetary systems known or suspected to exist around the 
massive evolved stars from our sample is most likely real.

\begin{acknowledgements}
      T.T.\ was supported by the International Max Plank Research School of Astronomy and Astrophysics - Heidelberg (IMPRS-HD)
and the Heidelberg Graduate School of Fundamental Physics (HGSFP). M.Z.\ acknowledges support by the European Research Council under the FP7 
Starting Grant agreement number 279347. A.R.\ has received financial support from the DFG under RE 
1664/9-2. This research has made use of the SIMBAD data base operated
at CDS, Strasbourg, France, the Vienna Atomic Line Database (VALD) at Institut f\"{u}r Astronomie, Wien, Austria, and
the HITRAN database at Harvard-Smithsonian Center for Astrophysics (CFA), Cambridge, MA, USA.
We thank the anonymous referee for the excellent comments that helped to improve this paper.

\end{acknowledgements}

\bibliographystyle{aa}

\bibliography{Trifonov_2015}

 % \clearpage

%\section{Appendix A:}

\begin{appendix} %First online appendix
 
 \setcounter{table}{0}
\renewcommand{\thetable}{A\arabic{table}}

 \setcounter{figure}{0}
\renewcommand{\thefigure}{A\arabic{figure}}

\begin{table*}{}        

\caption{Radial velocities measured with CRIRES}   
\resizebox{0.49\textheight}{!}{\begin{minipage}{\textwidth}

\label{table:RVs} 
 \begin{adjustwidth}{0.10cm}{}
 
 \begin{tabular}{ ccccc p{5.5mm} ccccc p{5.5mm} ccccc }
\cline{1-5}\cline{1-5}\cline{7-11}\cline{7-11}  \cline{13-17}\cline{13-17} 
\noalign{\vskip 0.5mm}
\cline{1-5}\cline{1-5}\cline{7-11}\cline{7-11}  \cline{13-17}\cline{13-17} 
\noalign{\vskip 0.5mm}

Target & & epoch [JD] & RV [km\,s$^{-1}$] &$\sigma$ [m\,s$^{-1}$]  && Target && epoch [JD] & RV [km\,s$^{-1}$]&$\sigma$ [m\,s$^{-1}$] && Target && epoch [JD] & RV [km\,s$^{-1}$]&$\sigma$ [m\,s$^{-1}$]    \\  
\cline{1-5}\cline{1-5}\cline{7-11}\cline{7-11} \cline{13-17}\cline{13-17} 
\noalign{\vskip 0.9mm}

HIP 5364        & & 2455853.844    & 11.623       & 19  &   &                    & & 2456288.592    & $-$34.193    & 29   &   &                   & & 2456453.802    & $-$23.342    & 15   \\  
                & & 2455854.616    & 11.588       & 40  &   &                    & & 2456326.559    & $-$34.333    & 55   &   &                   & & 2456496.664    & $-$23.374    & 17   \\ 
                & & 2456113.863    & 11.697       & 20  &   &                    & & 2456406.504    & $-$34.328    & 47   &   &                   & & 2456503.652    & $-$23.376    & 9    \\  \cline{7-11}\cline{13-17}\noalign{\vskip 0.7mm}
                & & 2456121.811    & 11.717       & 22  &   &    HIP 37826       & & 2455866.855    & 3.706         & 17  &   &   HIP 80693       & & 2455990.888    & 7.561         & 13   \\ 
                & & 2456139.747    & 11.662       & 12  &   &                    & & 2455896.833    & 3.728         & 15  &   &                   & & 2456015.773    & 7.956         & 11   \\ 
                & & 2456239.563    & 11.623       & 41  &   &                    & & 2455918.718    & 3.712         & 50  &   &                   & & 2456025.680    & 7.820         & 21   \\ 
                & & 2456441.910    & 11.774       & 33  &   &                    & & 2455994.554    & 3.759         & 19  &   &                   & & 2456025.676    & 7.832         & 24   \\ 
                & & 2456469.917    & 11.801       & 26  &   &                    & & 2456021.584    & 3.762         & 31  &   &                   & & 2456121.559    & 7.813         & 50   \\ 
                & & 2456494.892    & 11.800       & 26  &   &                    & & 2456288.709    & 3.623         & 20  &   &                   & & 2456406.714    & 7.826         & 13   \\  \cline{1-5}\noalign{\vskip 0.7mm}
HIP 19011       & & 2455854.736    & 26.864       & 6   &   &                    & & 2456326.567    & 3.647         & 15  &   &                   & & 2456459.675    & 7.956         & 19   \\
                & & 2455876.638    & 26.854       & 8   &   &                    & & 2456406.514    & 3.651         & 32  &   &                   & & 2456511.625    & 8.063         & 9    \\ \cline{7-11}\cline{13-17}\noalign{\vskip 0.7mm}
                & & 2456141.856    & 27.007       & 5   &   &    HIP 38253       & & 2455859.837    & $-$3.777     & 9    &   &    HIP 84671      & & 2455994.793    & 38.930        & 8    \\ 
                & & 2456147.907    & 27.020       & 7   &   &                    & & 2455896.774    & $-$3.822     & 20   &   &                   & & 2456018.858    & 39.056        & 10   \\ 
                & & 2456244.602    & 26.923       & 12  &   &                    & & 2455918.723    & $-$3.840     & 24   &   &                   & & 2456105.639    & 39.164        & 50   \\ 
                & & 2456287.561    & 26.917       & 23  &   &                    & & 2455973.736    & $-$4.039     & 17   &   &                   & & 2456121.670    & 39.315        & 145  \\  \cline{1-5}\noalign{\vskip 0.7mm}
HIP 20889       & & 2455854.744    & 38.407       & 14  &   &                    & & 2456021.597    & $-$3.936     & 21   &   &                   & & 2456148.674    & 39.137        & 21   \\ \cline{13-17}\noalign{\vskip 0.7mm}
                & & 2455900.760    & 38.392       & 24  &   &                    & & 2456244.742    & $-$4.311     & 12   &   &   HIP 88048       & & 2455997.887    & 12.901        & 23   \\ 
                & & 2456147.921    & 38.620       & 15  &   &                    & & 2456288.586    & $-$3.759     & 18   &   &                   & & 2456018.852    & 12.912        & 6    \\ 
                & & 2456168.832    & 38.604       & 14  &   &                    & & 2456326.571    & $-$3.851     & 8    &   &                   & & 2456105.635    & 13.048        & 49   \\ \cline{7-11}\noalign{\vskip 0.7mm}
                & & 2456244.649    & 38.618       & 27  &   &    HIP 39177       & & 2455859.843    & 39.473        & 13  &   &                   & & 2456113.599    & 13.068        & 28   \\ 
                & & 2456288.569    & 38.571       & 20  &   &                    & & 2455896.768    & 39.511        & 5   &   &                   & & 2456121.780    & 13.093        & 29   \\ 
                & & 2456495.931    & 38.438       & 13  &   &                    & & 2455918.731    & 39.502        & 7   &   &                   & & 2456148.679    & 13.206        & 26   \\ 
                & & 2456552.779    & 38.499       & 9   &   &                    & & 2455973.728    & 39.390        & 9   &   &                   & & 2456453.870    & 12.966        & 11   \\  \cline{1-5}\noalign{\vskip 0.7mm}
HIP 23015       & & 2455854.777    & 16.997       & 9   &   &                    & & 2456021.589    & 39.500        & 13  &   &                   & & 2456496.673    & 12.882        & 20   \\ 
                & & 2455900.755    & 16.728       & 7   &   &                    & & 2456288.701    & 39.254        & 8   &   &                   & & 2456497.759    & 12.925        & 16   \\ 
                & & 2456168.898    & 17.143       & 17  &   &                    & & 2456326.578    & 39.264        & 4   &   &                   & & 2456497.767    & 12.920        & 19    \\ \cline{7-11}\cline{13-17}\noalign{\vskip 0.7mm}
                & & 2456169.900    & 17.222       & 31  &   &    HIP 60202       & & 2455932.826    & 43.196        & 10  &   &   HIP 91004       & & 2455855.571    & $-$3.084     & 28    \\ 
                & & 2456244.738    & 16.933       & 16  &   &                    & & 2455973.744    & 43.302        & 10  &   &                   & & 2455997.891    & $-$3.574     & 9    \\ 
                & & 2456288.574    & 16.320       & 17  &   &                    & & 2456024.594    & 43.431        & 6   &   &                   & & 2456114.775    & $-$3.654     & 3    \\  \cline{1-5}\noalign{\vskip 0.7mm}
HIP 31592       & & 2455854.862    & 2.606        & 29  &   &                    & & 2456067.482    & 43.650        & 37  &   &                   & & 2456119.832    & $-$3.580     & 41   \\ 
                & & 2455876.701    & 2.564        & 18  &   &                    & & 2456293.849    & 43.255        & 13  &   &                   & & 2456119.846    & $-$3.538     & 23    \\ 
                & & 2455918.706    & 2.575        & 53  &   &                    & & 2456407.498    & 43.586        & 15  &   &                   & & 2456141.792    & $-$3.416     & 25   \\ 
                & & 2456021.552    & 2.576        & 19  &   &                    & & 2456458.561    & 43.356        & 37  &   &                   & & 2456239.543    & $-$2.432     & 4    \\ \cline{7-11}\cline{13-17}\noalign{\vskip 0.7mm}
                & & 2456039.484    & 2.601        & 22  &   &     HIP 73133      & & 2455937.846    & 17.637        & 6   &   &   HIP 100587      & & 2455860.541    & $-$15.468    & 4   \\ 
                & & 2456244.653    & 2.652        & 59  &   &                    & & 2455994.777    & 17.749        & 12  &   &                   & & 2456113.859    & $-$15.268    & 6   \\ 
                & & 2456288.579    & 2.630        & 18  &   &                    & & 2456022.713    & 17.744        & 15  &   &                   & & 2456119.838    & $-$15.278    & 11    \\ 
                & & 2456326.544    & 2.702        & 22  &   &                    & & 2456067.488    & 17.808        & 11  &   &                   & & 2456144.750    & $-$15.207    & 50    \\ 
                & & 2456406.497    & 2.695        & 24  &   &                    & & 2456406.698    & 18.099        & 34  &   &                   & & 2456405.903    & $-$14.705    & 11 \\  \cline{1-5}\noalign{\vskip 0.7mm}
HIP 34693       & & 2455859.855    & 22.027       & 11  &   &                    & & 2456458.572    & 18.196        & 13  &   &                   & & 2456440.905    & $-$14.680    & 20   \\ \cline{7-11}\noalign{\vskip 0.7mm}
                & & 2455895.863    & 22.222       & 23  &   &     HIP 74732      & & 2455939.840    & $-$9.514     & 10   &   &                   & & 2456466.747    & $-$14.607    & 37 \\ \cline{13-17}\noalign{\vskip 0.7mm}
                & & 2455918.699    & 22.270       & 6   &   &                    & & 2455994.784    & $-$9.493     & 13   &   &   HIP 114855      & & 2455854.611    & $-$25.721    & 43    \\  
                & & 2455994.537    & 21.845       & 14  &   &                    & & 2456024.647    & $-$9.526     & 6    &   &                   & & 2456112.928    & $-$25.563    & 17  \\
                & & 2456039.510    & 21.592       & 25  &   &                    & & 2456067.495    & $-$9.601     & 9    &   &                   & & 2456119.880    & $-$25.576    & 43   \\ 
                & & 2456244.866    & 22.114       & 13  &   &                    & & 2456406.729    & $-$9.466     & 10   &   &                   & & 2456144.774    & $-$25.691    & 26    \\ 
                & & 2456288.696    & 21.846       & 14  &   &                    & & 2456453.796    & $-$9.350     & 13   &   &                   & & 2456239.555    & $-$25.773    & 33   \\ 
                & & 2456326.553    & 21.644       & 12  &   &                    & & 2456494.649    & $-$9.323     & 18   &   &                   & & 2456453.898    & $-$25.617    & 86   \\  \cline{1-5}\cline{7-11}\noalign{\vskip 0.7mm} 
HIP 36616       & & 2455859.849    & $-$35.057    & 17  &   &    HIP 79540       & & 2456015.780    & $-$23.336    & 14   &   &                   & & 2456476.910    & $-$25.545    & 73   \\   
                & & 2455896.842    & $-$34.848    & 72  &   &                    & & 2456105.644    & $-$23.399    & 22   &   &                   & & 2456511.637    & $-$25.691    & 55    \\ 
                & & 2455918.683    & $-$34.780    & 33  &   &                    & & 2456113.594    & $-$23.406    & 18   &   &                   & &     &      &    \\ 
                & & 2455994.545    & $-$34.800    & 18  &   &                    & & 2456121.644    & $-$23.459    & 25   &   &                   & &     &      &    \\ 
                & & 2456021.576    & $-$34.910    & 49  &   &                    & & 2456406.706    & $-$23.391    & 22   &   &                   & &     &      &    \\ 
\cline{1-5}\cline{1-5}\cline{7-11}\cline{7-11}  \cline{13-17}\cline{13-17} 
 
\end{tabular}
\end{adjustwidth}
\end{minipage}}
\end{table*}

 \begin{figure*}[ht]

\includegraphics[width=6.3in]{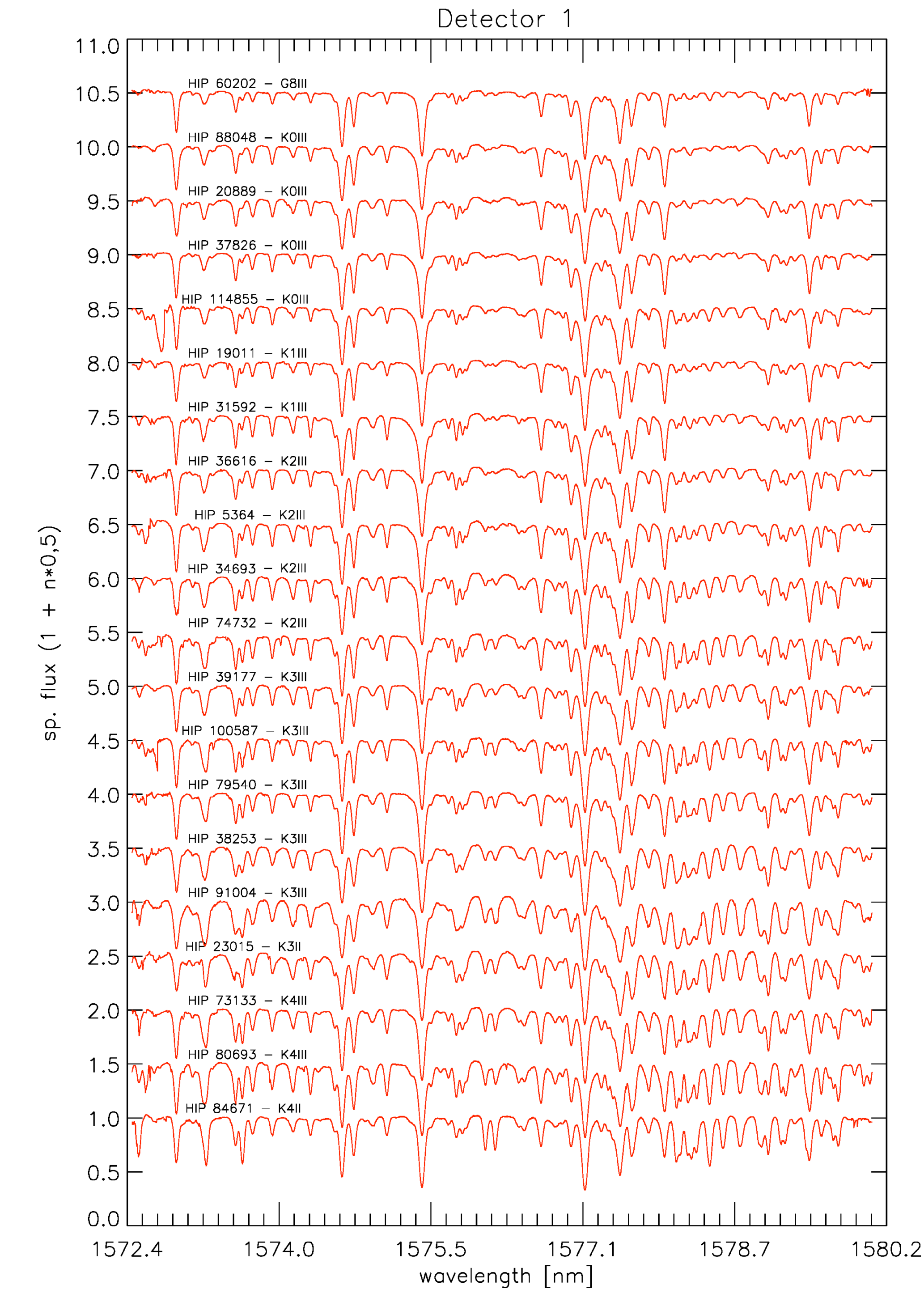}
\caption{Median-combined spectra from detector 1 for all observed targets, shifted by +0.5 in ascending order
starting from K4\,III giants at the bottom to earlier spectral type stars toward the top. 
Telluric lines were removed. 
Obviously late G and early K giants have a smaller number of deep stellar lines in 
this wavelength region than later K giants. 
} 

\label{FigGam:Mask_Median1}
\end{figure*}

\begin{figure*}[ht]

\includegraphics[width=6.3in]{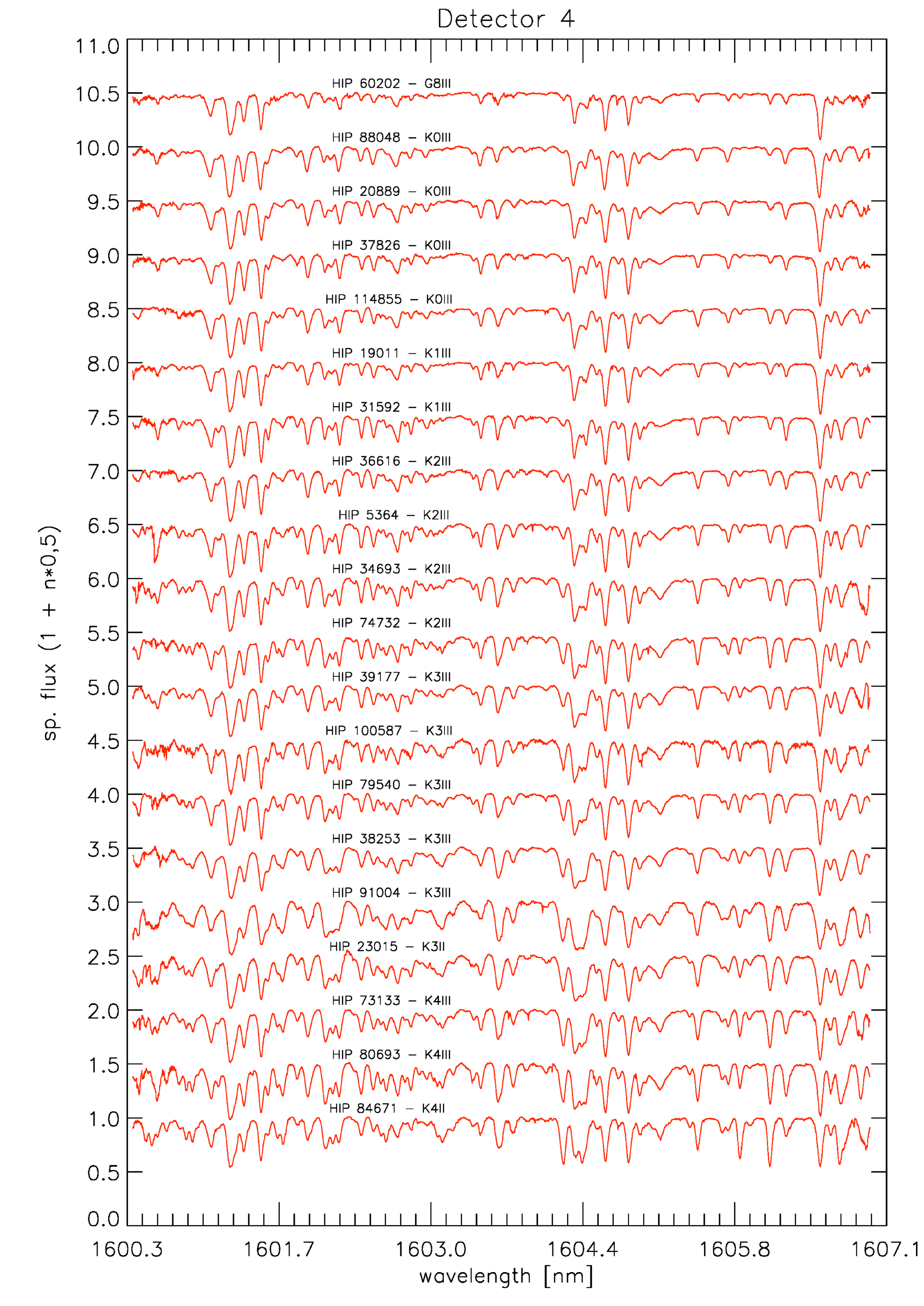}
\caption{Same as Fig.~\ref{FigGam:Mask_Median1}, except for detector 4. 
Detector 4 has in general lower S/N when compared to detector 1.
}
\label{FigGam:Mask_Median4}
\end{figure*}

\end{appendix}

\end{document}